\documentclass{article}
\usepackage{graphicx} 
\usepackage{array}
\usepackage{authblk}
\usepackage{amsmath,amsthm,amssymb,amsfonts,mathtools,empheq}
\usepackage{titlesec}
\usepackage{bbm}
\usepackage{booktabs}
\usepackage{url}
\usepackage{subfig}
\usepackage[colorlinks]{hyperref}
\usepackage[nameinlink,noabbrev]{cleveref}
\usepackage[labelfont=bf]{caption}
\usepackage[a4paper, total={6in, 9in}]{geometry}

\allowdisplaybreaks

\titleformat{\paragraph}
{\normalfont\normalsize\bfseries}{\theparagraph}{1em}{}
\titlespacing*{\paragraph}
{0pt}{3.25ex plus 1ex minus .2ex}{1.5ex plus .2ex}

\title{Improving Capital Efficiency and Impermanent Loss: Multi-Token Proactive Market Maker}

\author[1,*]{Wayne Chen}
\author[2]{Songwei Chen}
\author[3]{Preston Rozwood}
\affil[1]{\textit{zc272@cornell.edu}}
\affil[2]{\textit{songweichen@gmail.com}}
\affil[3]{\textit{prestonrozwood@gmail.com}}
\affil[*]{\textit{corresponding author}}
\date{January 3rd, 2025}

\begin{document}
\maketitle

\section*{Abstract}

Current approaches to the cryptocurrency automated market makers result in poor impermanent loss and capital efficiency. We analyze the mechanics underlying DODO Exchange's proactive market maker (PMM) to probe for solutions to these issues, leading to our key insight of multi-token trading pools. We explore this paradigm primarily through the construction of a generalization of PMM, the multi-token token proactive market maker (MPMM). We show via simulations that MPMM has better impermanent loss and capital efficiency than comparable market makers under a variety of market scenarios. We also test multi-token generalizations of other common 2-token pool market makers. Overall, this work demonstrates several advantages of multi-token pools and introduces a novel multi-token pool market maker.

\section*{Keywords}
blockchain, decentralized finance, decentralized exchanges, capital efficiency

\section{Introduction}

Automated Market Makers (AMMs) have evolved significantly to address issues such as capital inefficiency, price impact, impermanent loss (IL), and arbitrage opportunities. Capital inefficiency, price impact, and arbitrage are all consequences of exchange rate swings stemming from swapping tokens and they negatively affect both liquidity providers (LPs) and users of AMMs. Capital inefficiency refers to the deviation of a transaction's exchange rate from the fair market price (determined by dividing tokens' prices in a traditional currency), price impact refers to the magnitude of an AMM exchange rate change following a transaction, and arbitrage refers to transacting across multiple trading platforms to capitalize on mismatched exchange rates. While IL does not affect traders directly, it does affect LPs who supply tokens to AMMs in exchange for a cut of the transaction fees. IL is the phenomenon where the AMM portfolio value degrades in comparison to a simple holding strategy over time. There have been a number of studies addressing these and related issues.
\\\\
Uniswap V2~\cite{Adams2020UniswapVC} is based on a constant product market making (CPMM) AMM, which maintains that the product of token balances in a pool always equals a constant. CPMM is an extension of a more naive constant sum market maker (CSMM) that simply always executes swaps at a fair market exchange rate. CSMM ensures no price impact, but allows arbitrageurs to drain one of the reserves if the off-chain reference price between the tokens is inaccurate~\cite{chainlink_amm_types}. Under CPMM, LPs deposit $X$ units of token A and $Y$ units of token B, where the ratio $Y/X$ is the implicit marginal price of an A token measured in B tokens and is typically reflective of the current fair market price due to arbitrageurs~\cite{uniswap_liquidity_docs}. The simplicity of the pricing rule gives rise to persistent arbitrage opportunities and consistently profitable front-running~\cite{flaws_of_cpmm}. Front-running occurs when traders capitalize on exchange rate movements caused by transactions within a new pending block of the blockchain. Furthermore, simulations and real-world analyses reveal that Uniswap V2 LPs receive inconsequential annualized returns of up to 1\% primarily due to competition between similarly resourced LPs. As expected, trading volume is directly correlated with increased LP profitability~\cite{uni_v2_analysis}.
\\\\
Uniswap V3~\cite{Adams2021UniswapVC} introduces the concept of concentrated liquidity to its CPMM, allowing LPs to allocate funds within specific price ranges. This design incentivizes concentrating liquidity in prices of high trading activity (typically those surrounding fair market prices), thus improving capital efficiency. However, providing liquidity in Uniswap V3 requires active management and acceptance of financial risks, as returns can vary widely depending on market conditions.
Consequently, Uniswap V3 has pushed away smaller, less sophisticated liquidity providers~\cite{uni_v3_risks}. IL has also been studied within Uniswap V3: the AMM's liquidity provision structure increases fee earnings, but exacerbates IL by narrowing trading ranges. In fact, in major Uniswap V3 pools, LPs often incur greater losses from IL than they earn in fees, leading to worse outcomes than simply holding their assets~\cite{uni_v3_il}.
\\\\
Both Uniswap V2 and V3's designs reveal a struggle in balancing needs of AMM users and LPs. A barebones CPMM simplifies liquidity provision but is capitally inefficient, leading to arbitrage and front-running. Sophisticated additions such as concentrated liquidity address capital efficiency to eliminate arbitrage and front-running, but are too complex and risky for LPs. Although several years old already, the Uniswap V2 and V3's designs are continually reused. Uniswap remains the largest exchange today by total volume locked. Raydium, the 2nd largest, has CPMM pools and order books~\cite{raydium}. Order books pair buy and sell requests, thus bypassing exchange rate issues and avoiding liquidity provision and its problems altogether, but are only effective with sufficient liquidity~\cite{order-book}. Meanwhile, the 3rd largest, Aerodrome is forked from Uniswap V3 \cite{aerodrome}.
\\\\
Proactive Market Maker (PMM), as implemented by DODO Exchange~\cite{dodo_docs}, innovates beyond the CPMM to mitigate impermanent loss and improve capital efficiency without introducing obscure functionality, therefore addressing both LPs and AMM users' needs. To improve capital efficiency, PMM introduces a parameter $k$ that flattens the exchange rate curve to match the fair market exchange rate, accessible via a price oracle, over a broad liquidity range surrounding an adjustable token balance ``equilibrium". The equilibrium is initially determined by LPs' deposit amounts and later adjustments minimize movements to the equilibrium, thus keeping pools in line with a holding strategy and minimizing impermanent loss. Additionally, LPs can supply liquidity using single token types, reducing barriers to participation.
\\\\
Our contributions include an analysis of PMM's mechanics leading to the key finding that \textit{increased liquidity boosts its capital efficiency and better prevents impermanent loss}. We explore this concept through MPMM, a multi-token generalization of PMM that aggregates all the pairwise trading pools into a single large pool. We compare MPMM and PMM against other multi-token and 2-token pool AMMs, finding that MPMM outperforms its counterparts in terms of capital efficiency and impermanent loss under a variety of market conditions.

\section{Methods}
\subsection{Preliminary: Proactive Market Maker}

Consider a token pool with 2 tokens. Let $B$ denote the current balance of the first token type and $P_B$ denote its true market price. Similarly, define $Q$ and $P_Q$ for the second token type. Let the price of one token with respect to the other be governed by an exchange rate function dependent on the balances of the 2 token types. Let $B_0$ and $Q_0$ denote special balances of the first and second token type such that the exchange rate at this state equals $\frac{P_B}{P_Q}$ or $\frac{P_Q}{P_B}$ (i.e. the fair market exchange rate). Finally, let $0 < k < 1$ be a parameter controlling the amount of deviation from the market exchange rates at pool balances far away from $B_0$ and $Q_0$. The PMM pool balance is described by the following piecewise function~\cite{dodo_docs}:

\begin{subequations}\label{eq:system}
\begin{empheq}[left=\empheqlbrace]{align}
  &Q = -\frac{P_B}{P_Q}(B - B_0)\left(1 - k + \frac{kB_0}{B} \right) + Q_0 \qquad 0 < B \leq B_0 \label{eq:system_1}
  \\
  & B = -\frac{P_Q}{P_B}(Q - Q_0)\left(1 - k + \frac{kQ_0}{Q} \right) + B_0 \qquad 0 < Q < Q_0 \label{eq:system_2}
\end{empheq}
\end{subequations}

The resulting curve has marginal exchange rates

\begin{equation}
\begin{split}\label{eq:marginals_short}
    \frac{\partial Q}{\partial B} &= \frac{P_B(B_0^2k - (k-1)B^2)}{P_Q B^2}\\
    \frac{\partial B}{\partial Q} &= \left(P_Q + \frac{P_Q(P_B B_0 (1 - 2k) + P_Q (Q_0 - Q))}{\sqrt{(P_B B_0 (1 - 2k) + P_Q (Q_0 - Q))^2 - 4 P_B^2 B_0^2 k (k-1)}} \right) \cdot (2P_B(k-1))^{-1}
\end{split} 
\end{equation}

when $0 < B \leq B_0$. When $0 < Q < Q_0$, the exchange rates are
\begin{equation}
\begin{split}\label{eq:marginals_long}
    \frac{\partial B}{\partial Q} &= \frac{P_Q(Q_0^2k - (k-1)Q^2)}{P_B Q^2}\\
    \frac{\partial Q}{\partial B} &= \left(P_B + \frac{P_B(P_Q Q_0 (1 - 2k) + P_B (B_0 - B))}{\sqrt{(P_Q Q_0 (1 - 2k) + P_B (B_0 - B))^2 - 4 P_Q^2 Q_0^2 k (k-1)}} \right) \cdot (2P_Q(k-1))^{-1}
\end{split}
\end{equation}

where $\frac{\partial Q}{\partial B}$ is the marginal exchange rate for receiving tokens of the second type for one unit of the first token type added to the pool and $\frac{\partial B}{\partial Q}$ is the marginal exchange rate for receiving the tokens of the first type for one unit of the second token type added to the pool. It is guaranteed that exactly one of $0 \leq B \leq B_0$ or $0 \leq Q \leq Q_0$ is always true. The derivations can be found in the \hyperref[sec:proof_theorem_1]{appendix}.

\subsubsection{Difficulty of Draining}
One important observation from \hyperref[eq:system]{1} is that \textit{it is impossible for either token type to be completely drained from a pool}. Specifically,

\begin{align*}
    \lim_{\frac{P_Q}{P_B} \rightarrow \infty} \frac{\partial Q}{\partial B} &= 0, \text{ if } Q < Q_0\\
    \lim_{\frac{P_B}{P_Q} \rightarrow \infty} \frac{\partial B}{\partial Q} &= 0, \text{ if } B < B_0\\
\end{align*}

The proof of these limits can be found in the \hyperref[sec:proof:2_1_1]{appendix}. This implies that attempting to mint large amounts of one token type (thus deflating its price) to buy the other token is impossible. A caveat is that since the PMM pricing curve uses $P_B$ and $P_Q$ provided from oracles, the oracles must quickly adapt to price fluctuations.

\subsubsection{Self Balancing Exchange Rates}\label{sec:2.1.2}
Note that:

\begin{align*}
    \lim_{Q \rightarrow Q_0} \frac{\partial B}{\partial Q} &= \frac{P_Q}{P_B}, \text{ if } Q < Q_0\\
    \lim_{B \rightarrow B_0} \frac{\partial Q}{\partial B} &= \frac{P_B}{P_Q}, \text{ if } B \leq B_0\\
    \frac{\partial B}{\partial Q} &> \frac{P_Q}{P_B}, \text{ if } Q < Q_0\\
    \frac{\partial Q}{\partial B} &> \frac{P_B}{P_Q}, \text{ if } B < B_0
\end{align*}

The proofs can be found in the \hyperref[sec:proof:2_1_2]{appendix}. The first two observations indicate that with token balances at the equilibrium $B_0$ and $Q_0$, there are no arbitrage opportunities (assuming $P_Q$ and $P_B$ are accurate). The second two shows that any other pool state is inclined to return to equilibrium due to the existence of profitable arbitrage. In summary, \textit{assuming the ratio $P_Q/P_B$ remains fixed, token balances will naturally return to equilibrium}. DODO sets their PMM $B_0$ and $Q_0$ equal to the token amounts deposited by LPs~\cite{dodo_docs} to ensure the pool tracks a holding strategy, thus reducing impermanent loss. Note that a similar notion of $B_0$ and $Q_0$ also implicitly exists within CPMMs which require LPs to deposit 2 token types in amounts reflecting their fair market exchange rate--however PMM makes these parameters settable, thus allowing provisioning of single token type.

\subsubsection{Combating Impermanent Loss with Excess Tokens} \label{sec:2.1.3}
During most times, the token balances would not be at the equilibrium point $B_0$ and $Q_0$ exactly equal to the count of LPs' provided tokens. During such times and when the exchange rate $P_Q/P_B$ has also shifted, PMM takes sets a new equilibrium that is ``as close to optimal" as possible.
\\
\\
Suppose prices have shifted from $P_B$ and $P_Q$ to $P_{B'}$ and $P_{Q'}$. Then, PMM derives the following new equilibrium points $B_{0'}$ and $Q_{0'}$

\begin{equation}\label{eq:equilibrium_short}
\begin{split}
    B_{0'} &= B + \frac{B}{2k} \left(\sqrt{1 + \frac{4k(Q - Q_0)}{\frac{P_{B'}}{P_{Q'}} B}} - 1 \right)\\
    Q_{0'} &= Q_0
\end{split}
\end{equation}

if $B < B_0$ and

\begin{equation}\label{eq:equilibrium_long}
\begin{split}
    Q_{0'} &= Q + \frac{Q}{2k} \left(\sqrt{1 + \frac{4k(B - B_0)}{\frac{P_{Q'}}{P_{B'}} Q}} - 1 \right)\\
    B_{0'} &= B_0
\end{split}
\end{equation}

if $Q < Q_0$ immediately before processing a transaction. Otherwise, the existing equilibrium is kept.
\\
\\
This mechanism utilizes the token type that is in excess (with respect to the original equilibrium) to \textit{recoup as much of the token type in shortage as possible while maintaining a valid PMM balance state curve}. More details regarding a proof on this point can be found in the \hyperref[sec:proof:2_1_3]{appendix}. 

\subsubsection{\textit{k} Balances Risk and Efficiency} \label{sec:2.1.4}
Though $k$ is never 0 or 1, we can study \hyperref[eq:system]{1} under these conditions to bound properties of the real curves. With $k=0$, the balances follow the curve

\begin{equation}\label{eq:csmm}
    Q + \frac{P_B}{P_Q}B = Q_0 + \frac{P_B}{P_Q}B_0
\end{equation}

and $k=1$ results in the piecewise function

\begin{subequations}\label{eq:cpmm}
\begin{empheq}[left=\empheqlbrace]{align}
  &Q = -\frac{P_B}{P_Q}(B - B_0)\frac{B_0}{B} + Q_0 \qquad 0 < B \leq B_0
  \\
  &B = -\frac{P_Q}{P_B}(Q - Q_0)\frac{Q_0}{Q} + B_0 \qquad 0 < Q < Q_0
\end{empheq}
\end{subequations}

\hyperref[eq:csmm]{6} is a variant of CSMM, while the \hyperref[eq:cpmm]{7} is a variant of CPMM (more details in the \hyperref[sec:proof_3_1_4]{appendix}). With $0 < k < 1$, many properties are a middle ground between those of the CSMM and CPMM. For example:

\begin{enumerate}
    \item Token draining: It is impossible to drain any token's reserve for CPMM and also for PMM as long as $k \neq 0$ since \hyperref[eq:system]{1} asymptotically approaches 0. However, \hyperref[eq:csmm]{6} is a line, so a token can be depleted in a CSMM pool.

    \item Changing exchange rates: CSMM maintains a constant exchange rate equal to the fair market exchange rate. All other AMM formulas in this paper have changing exchange rates and a unique equilibrium point $B_0$ and $Q_0$ (see \hyperref[sec:2.1.2]{2.1.2}) where the exchange rate matches the fair market exchange rate. For PMM, a lower $k$ results in less exchange rate fluctuation near the equilibrium token balance: in \hyperref[eq:marginals_short]{2} and \hyperref[eq:marginals_long]{3}, $\frac{\partial Q}{\partial B} \rightarrow \frac{P_B}{P_Q}$ as $k \rightarrow 0$ for fixed $B$ and $\frac{\partial B}{\partial Q} \rightarrow \frac{P_Q}{P_B}$ as $k \rightarrow 0$ for fixed $Q$ (see the \hyperref[sec:proof_small_k]{appendix} for details). There are several notable effects of this:

    \begin{enumerate}
        \item A pool is more capital efficient with lower $k$, since the exchange rate does not deviate significantly from the fair market rate away from the equilibrium balance. In fact, CSMM, with $k=0$, achieves perfect capital efficiency since the exchange rate always equals the market rate.
        \item By similar logic, there is lower price impact with lower $k$ (and CSMM has no price impact).
        \item By similar logic, arbitrage and front-running profits are significantly reduced with a lower $k$.
    \end{enumerate}

    \item Ability to combat impermanent loss: CSMM does not have any ability to combat impermanent loss. Its trades always occur at market rates and there are no ways to use exchange rates to favor certain token balances. In contrast, CPMM and PMM will always have a unique equilibrium balance reflective of LPs' deposits where the exchange rate matches the market rate. However, increasing $k$ in PMM makes pools more resilient. This can be observed by noting that $k$ increases the calculated new equilibrium balance in \hyperref[eq:equilibrium_short]{4} and \hyperref[eq:equilibrium_long]{5} for the depleted token type. A high $k$ also makes swapping more expensive~\cite{dodo_docs}, giving pool more resources to recoup the token type that was depleted.
\end{enumerate}

\subsection{From Increased Liquidity to Multi-Token Proactive Market Maker Model}\label{sec:2.2}
A PMM with increased liquidity is less prone to exchange rate deviating significantly from the market rate when the pool is away from the equilibrium balance $B_0$ and $Q_0$ (thus improving protection against arbitrage/front-runners and boosting capital efficiency). Explicitly, in \hyperref[eq:marginals_short]{2} and \hyperref[eq:marginals_long]{3}, $\frac{\partial Q}{\partial B} \rightarrow \frac{P_B}{P_Q}$ and $\frac{\partial B}{\partial Q} \rightarrow \frac{P_Q}{P_B}$ with increased $B_0$ and $Q_0$ (see \hyperref[sec:liquidity_boosts_cap_eff]{appendix} for details).
\\\\
Increasing liquidity also boosts the effectiveness of the PMM recovery procedure described by \hyperref[eq:equilibrium_short]{4} and \hyperref[eq:equilibrium_short]{5}. Concretely, $B_{0'}$ and $Q_{0'}$ are able to recover closer to $B_0$ and $Q_0$ (see \hyperref[sec:liquidity_helps_imp_loss]{appendix} for details) once the exchange rate $P_Q/P_B$ has shifted, thus better aligning the pool with a holding strategy and increasing protection against impermanent loss.
\\\\
A simple way to build larger liquidity AMM pools is combining traditional 2-token pools into a single large multi-token pool. For example, one can imagine merging all the Uniswap pools with USDC to create a giant pool with significantly more USDC. Beyond the benefits for capital efficiency and impermanent loss, multi-token pools simplify swapping between lesser established tokens which typically do not have a shared trading pools in an AMM. The predominant solution for such swaps rely on complex order routing systems to find optimal exchange rates spanning multiple trading pairs~\cite{balancer} that require more gas fees.
\\\\
While generalizing CPMM to MCPMM is mathematically simple (and has already been done by ~\cite{balancer}), PMM's piecewise state curve is nontrivial to generalize. For simplicity, our MPMM is a PMM that allows swapping between arbitrary pairs of tokens (in the AMM), that derives new equilibriums $B_{0'}$ and $Q_{0'}$ for the swapped token pair differently than in \hyperref[sec:3.1.3]{3.1.3} by greedily optimizing MPMM to reduce the following before each trade:

\begin{equation}\label{eq:optimization}
    \left(1 - \frac{B_{0'}}{B_R} \right)^2 + \left(1 - \frac{Q_{0'}}{Q_R} \right)^2
\end{equation}

where $B_R$ is the amount of the 1st token type deposited by liquidity providers, and $Q_R$ the amount of the 2nd, subject to the constraint

$$B_{0'} = B + \frac{B}{2k} \left(\sqrt{1 + \frac{4k(Q - Q_{0'})}{\frac{P_B}{P_Q} B}} - 1 \right)$$

if $\frac{B_0}{B_R} < \frac{Q_0}{Q_R}$ or the constraint

$$Q_{0'} = Q + \frac{Q}{2k} \left(\sqrt{1 + \frac{4k(B - B_{0'})}{\frac{P_Q}{P_B} Q}} - 1 \right)$$

otherwise. Here, $k$ can be calculated a variety of methods, for example by first assigning to each token a $k$ value then taking an average. A closed form solution for this problem exists and was found using Wolfram Alpha, though it is too long to write out\footnote[1]{The full expression is displayed in the \href{https://github.com/Frutto-Group/multi-token-proactive-market-maker/blob/main/mpmm.py}{code} in the $\textit{\_\_argMin}$ function}.
\\
\\
\hyperref[eq:optimization]{8} aims to reduce a weighted ``distance" between the new equilibrium and the ideal equilibrium that tracks LPs' deposit amounts while maintaining a valid PMM-style curve. Furthermore, the procedure ensures that when traffic only involves 2 tokens, MPMM and PMM pools evolve identically.

\section{Simulations}

We benchmark PMM, CSMM, CPMM, and their multi-token generalizations (MPMM, MCSMM, MCPMM, respectively). For MCSMM and CSMM, all swaps occur at market exchange rates, and those that use up more tokens than available in the pool are not performed. For MCPMM, all swaps maintain that the product of all token balances in the pool would always equal a constant.
\\\\
We measure 2 statistics: capital efficiency and impermanent loss. Capital efficiency is calculated as

$$\frac{\Delta I / \Delta O}{P_O / P_I}$$

where $P_O$ and $P_I$ are the real world prices of the output and input token types, respectively, and $\Delta O$ and $\Delta I$ are the amount of tokens outputted from and inputted to the AMM, respectively. We use a generalization of the impermanent loss as described in~\cite{uni_v3_il} and calculate the statistic as

$$\frac{V_t - H_t}{H_0}$$

where $V_t$ is the value of a pool at timestep $t$ in a simulation, $H_t$ is the value of a hypothetical portfolio that only holds the pool's initial tokens until timestep $t$, and $H_0$ is the initial value of the pool.
\\\\
We calculate impermanent loss differently for different AMMs. For CSMM, MCSMM, PMM, and MPMM, LPs can provide single token types to pools, while CPMM and MCPMM AMMs require multiple token types. So, for CPMM and MCPMM, a 2-token and the single multi-token pool, respectively, is considered as 1 ``pool". For the remainder 2-token (multi-token) AMMs, each token within each 2-token (multi-token) pool is considered a ``pool". This is a reasonable since impermanent loss is associated with the withdrawal value of LPs' deposits, so the statistic should take into account provisioning mechanics.

\subsection{Real World Price Simulations}
\begin{figure*}[t!]
    \subfloat[CPMM variants]{%
        \includegraphics[width=.33\linewidth]{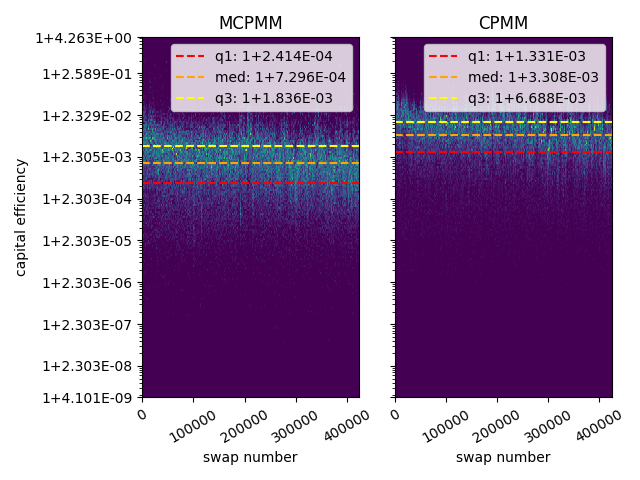}%
        \label{subfig:cpmm_cap_eff}%
    }\hfill
    \subfloat[PMM, k=0.05 variants]{%
        \includegraphics[width=.33\linewidth]{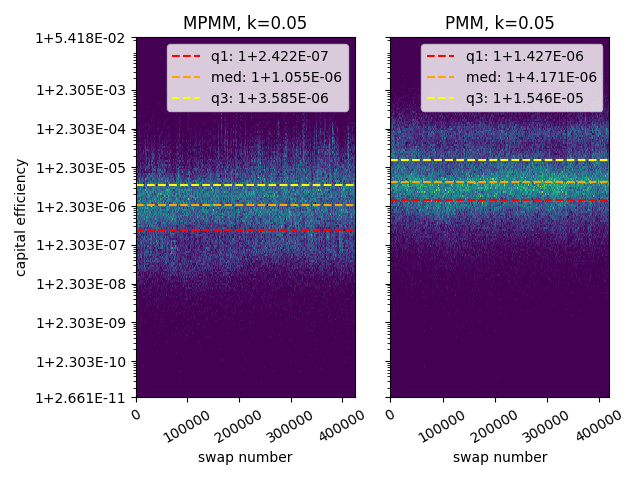}%
        \label{subfig:pmm_005_cap_eff}%
    }\hfill
    \subfloat[PMM, k=0.25 variants]{%
        \includegraphics[width=.33\linewidth]{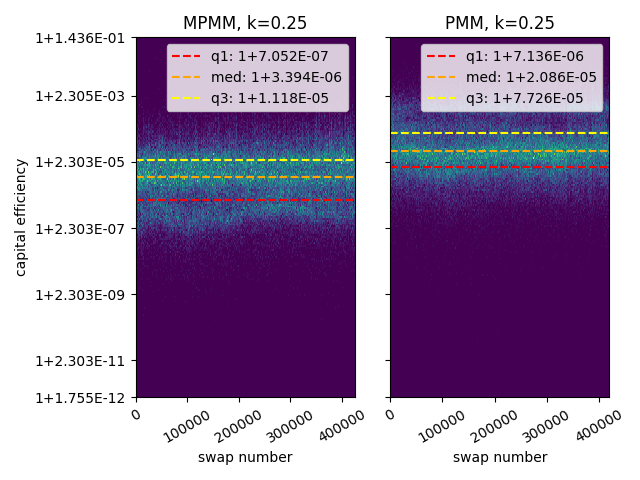}%
        \label{subfig:pmm_025_cap_eff}%
    }\\
    \subfloat[CSMM variants]{%
        \includegraphics[width=.33\linewidth]{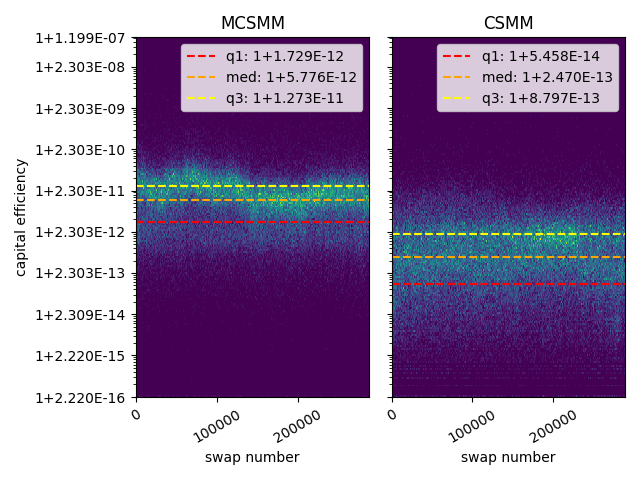}%
        \label{subfig:csmm_cap_eff}%
    }\hfill
    \subfloat[PMM, k=0.5 variants]{%
        \includegraphics[width=.33\linewidth]{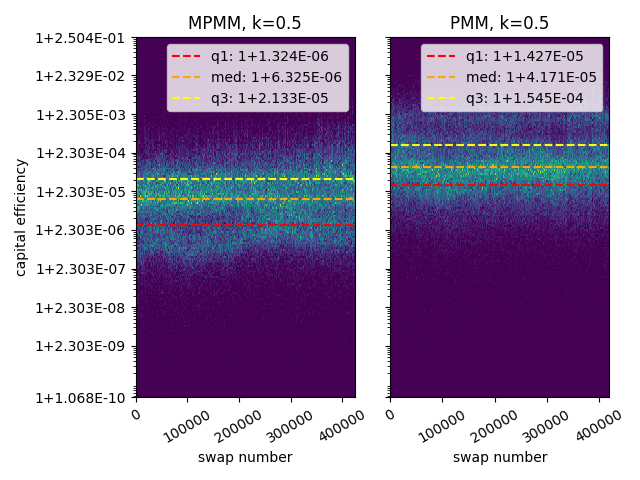}%
        \label{subfig:pmm_050_cap_eff}%
    }\hfill
    \subfloat[PMM, k=0.75 variants]{%
        \includegraphics[width=.33\linewidth]{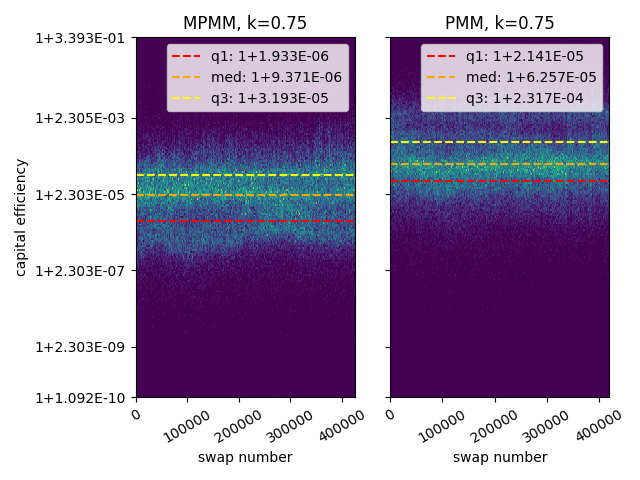}%
        \label{subfig:pmm_075_cap_eff}%
    }
    \caption{\textbf{Capital efficiency of AMMs over course of real world price simulation:} We chart the capital efficiency of swaps more costly than the fair market rate. Dashed lines indicate the 1st, 2nd, and 3rd quartiles. Note the different scaling for different plots.}
    \label{fig:normal_cap_eff}
\end{figure*}

\begin{figure*}[t!]
    \subfloat[CPMM Variants]{%
        \includegraphics[width=.33\linewidth]{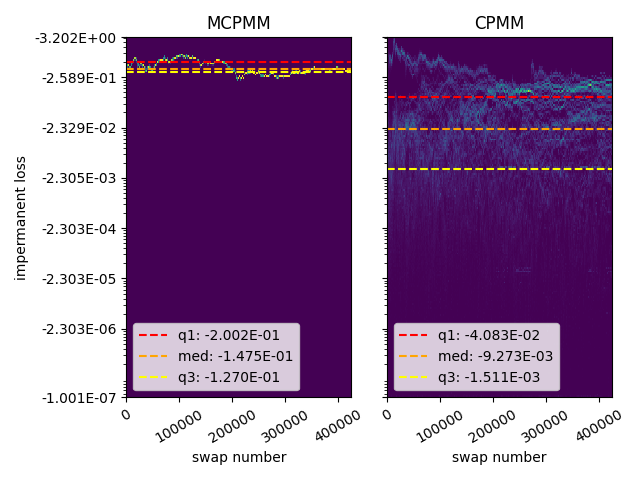}%
        \label{subfig:cpmm_imp_loss}%
    }\hfill
    \subfloat[PMM, k=0.05 variants]{%
        \includegraphics[width=.33\linewidth]{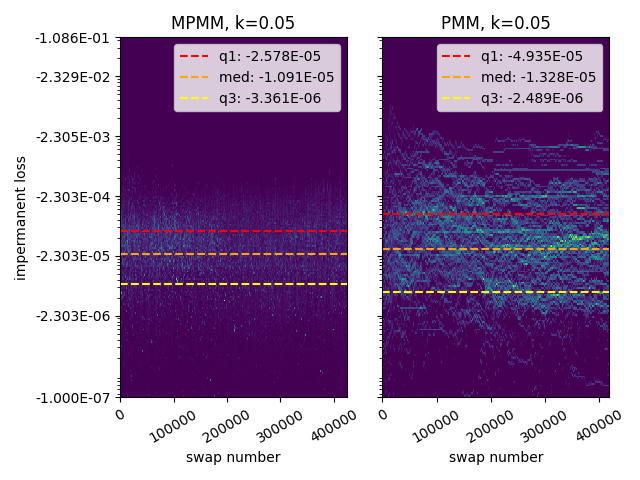}%
        \label{subfig:pmm_005_imp_loss}%
    }\hfill
    \subfloat[PMM, k=0.25 variants]{%
        \includegraphics[width=.33\linewidth]{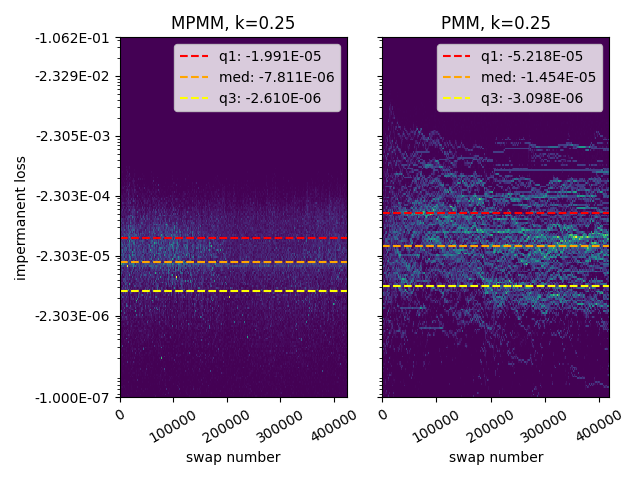}%
        \label{subfig:pmm_025_imp_loss}%
    }\\
    \subfloat[CSMM Variants]{%
        \includegraphics[width=.33\linewidth]{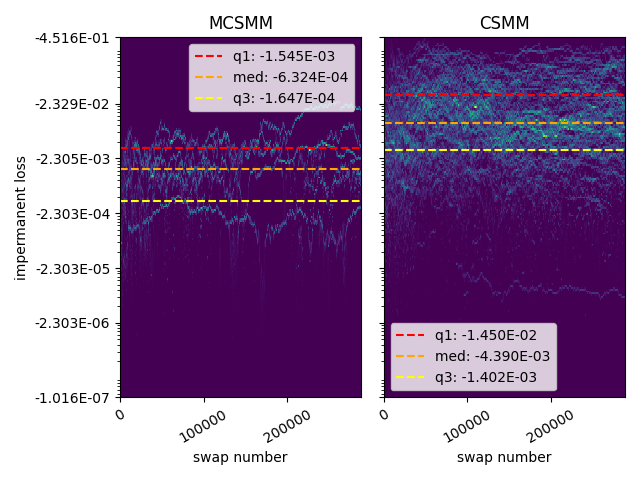}%
        \label{subfig:csmm_imp_loss}%
    }\hfill
    \subfloat[PMM, k=0.5 variants]{%
        \includegraphics[width=.33\linewidth]{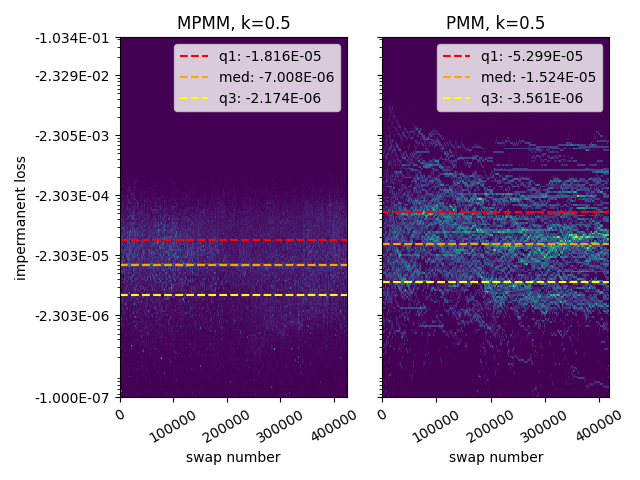}%
        \label{subfig:pmm_050_imp_loss}%
    }\hfill
    \subfloat[PMM, k=0.75 variants]{%
        \includegraphics[width=.33\linewidth]{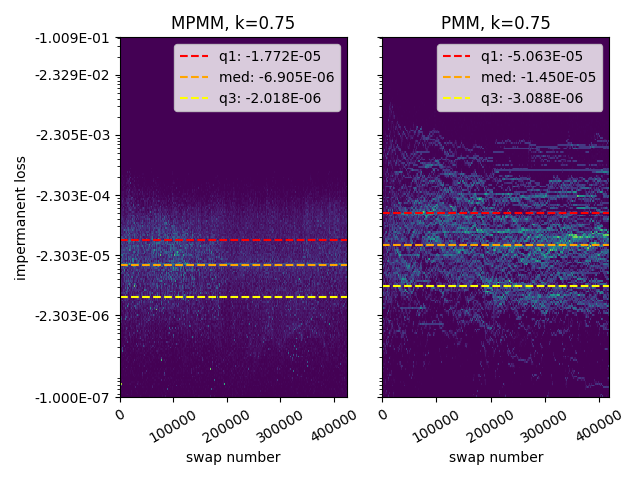}%
        \label{subfig:pmm_075_imp_loss}%
    }
    \caption{\textbf{Impermanent loss of AMMs over course of real world price simulation:} We chart the impermanent loss of pools following swaps where a loss has occurred.}
    \label{fig:normal_imp_loss}
\end{figure*}

We use hourly historical price data from a Kaggle dataset for 14 token types \cite{kaggle} spanning from April 8th, 2021 to September 10th, 2023\footnote{Overall in this period, the crypto market experienced a peak around November 2021 followed by a dropoff until June 2022, and then a period of stagnation until the simulation's end in September 2023.} and daily per-token transaction volumes from CoinGecko to generate swapping traffic. We estimate hourly per-token transaction volumes by linearly interpolating volumes from consecutive days. Then, for each hour in each day of the simulation, we randomly sample token pairs to swap 20\footnote{We note 20 is several orders of magnitude off from real life exchanges' traffics (using daily transaction volumes and average transaction sizes from Uniswap suggests this number should be in the 1000's), but 20 was chosen to ensure simulations completed within a reasonable amount of time.} times, with sampling weighted by the transaction volume of tokens that hour. The input dollar amount is sampled from a normal distribution with mean \$10,000\footnote{This is estimated using the first figure from \cite{dollar_size} (specifically checking values around April 8th, 2021).}.
\\\\
In multi-token market makers, the 14 tokens are combined into a single pool, while 91 2-token pools (1 for each unique combination of 2 tokens) are generated for the 2-token market makers. Single token pools are initialized to be worth 1\% of the token's market cap\footnote{1\% is roughly the right order of magnitude of a large exchange such as Uniswap.}. For 2-token pools, say of types $t_1$ and $t_2$, we set the total pool value of each token to be \textit{market cap}($t_1$) $\cdot$ \textit{market cap}($t_2$) / \textit{market cap}(all tokens)\footnote{This simplifying choice is made due to the difficulty of sourcing historical data for 91 token pools.}
\\\\
In \hyperref[fig:normal_cap_eff]{1}, we plot the capital efficiency of swaps with $\frac{\Delta I / \Delta O}{P_O / P_I} > 1$ (those more costly than otherwise possible given fair market prices) throughout the duration of the simulation. Overall, we see multi-token AMMs are more capital efficient than their 2-token pool variants (e.g. $1+7.296\mathrm{e}{-4}$ vs $1+3.308\mathrm{e}{-3}$ for MCPMM vs CPMM  and $1+1.055\mathrm{e}{-6}$ vs $1+4.171\mathrm{e}{-5}$ for MPMM vs PMM, $k = 0.5$, median capital efficiency), thus confirming our hypothesis that increasing liquidity helps capital efficiency. We also see that PMM and MPMM are significantly more efficient than CPMM and MCPMM but not as efficient as CSMM and MCSMM. The figure actually undersells CSMM and MCSMM, which always transacts at the market one, thus resulting in a capital efficiency of 1 always (we believe the plotted points are due to floating point errors). However, CSMM and MCSMM's results should mostly be ignored: they are never employed in real life due to their sensitivity to any pricing oracle inaccuracy will result in significant pool draining. We also see that increasing $k$ results in capital inefficiency for both PMM and MPMM (e.g. $1+1.055\mathrm{e}{-6} \rightarrow 1+9.371\mathrm{e}{-6}$ and $1+4.171\mathrm{e}{-6} \rightarrow 1+6.257\mathrm{e}{-5}$ median capital efficiency for MPMM and PMM going from $k=0.05$ to $k=0.75$).
\\\\
In \hyperref[fig:normal_imp_loss]{2}, we plot the impermanent loss in token pools when losses have occurred (i.e. negative values of $\frac{V_t - H_t}{H_0}$). Overall, we see that CPMM, CMPMM, CSMM, and MCSMM all suffer significantly more than PMM and MPMM. In fact, CSMM and MCSMM refuse to accept transactions midway through the simulation due to depleted token pools. Anther noteworthy comparison is that MCPMM performs worse than CPMM while MPMM outperforms PMM across a range of $k$ values (MPMM consistently has a lower median impermanent loss that is less volatile compared to PMM). We believe this is evidence for the point in \hyperref[sec:2.2]{2.2} regarding how the corrective mechanism in \hyperref[sec:2.1.3]{2.1.3} updates $B_0 \rightarrow B_{0'}$ and $Q_0 \rightarrow Q_{0'}$ less (proportional to $B_0$ and $Q_0$) when given more liquidity.

\subsection{Bull Market Simulation}
\begin{figure*}[t!]
    \subfloat[CPMM variants]{%
        \includegraphics[width=.33\linewidth]{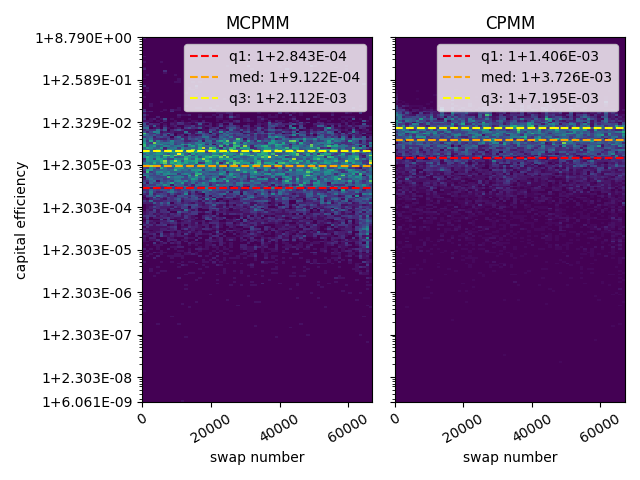}%
        \label{subfig:cpmm_cap_eff}%
    }\hfill
    \subfloat[PMM, k=0.05 variants]{%
        \includegraphics[width=.33\linewidth]{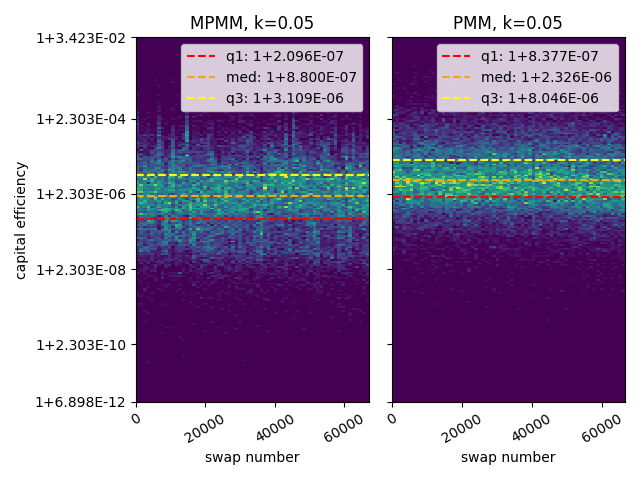}%
        \label{subfig:pmm_005_cap_eff}%
    }\hfill
    \subfloat[PMM, k=0.25 variants]{%
        \includegraphics[width=.33\linewidth]{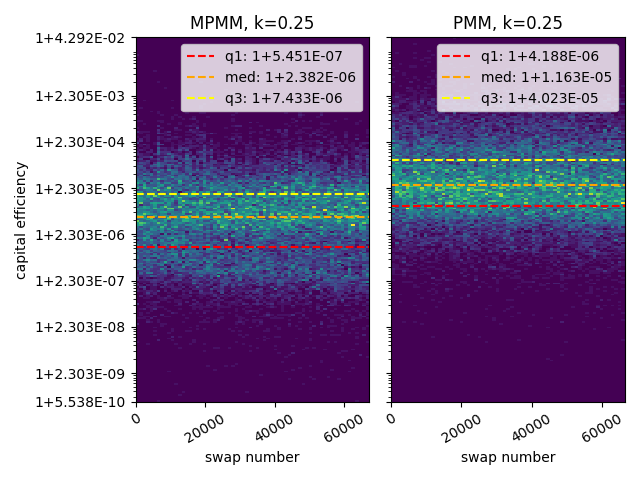}%
        \label{subfig:pmm_025_cap_eff}%
    }\\
    \subfloat[CSMM variants]{%
        \includegraphics[width=.33\linewidth]{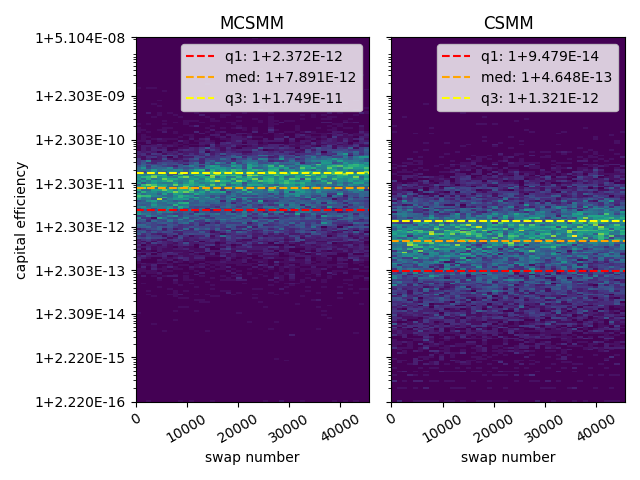}%
        \label{subfig:csmm_cap_eff}%
    }\hfill
    \subfloat[PMM, k=0.5 variants]{%
        \includegraphics[width=.33\linewidth]{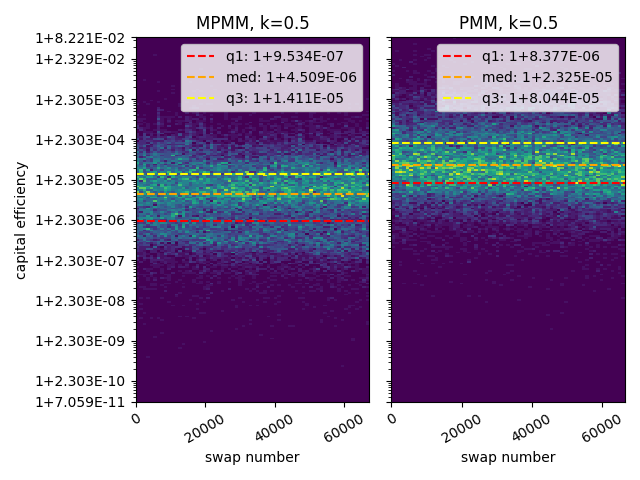}%
        \label{subfig:pmm_050_cap_eff}%
    }\hfill
    \subfloat[PMM, k=0.75 variants]{%
        \includegraphics[width=.33\linewidth]{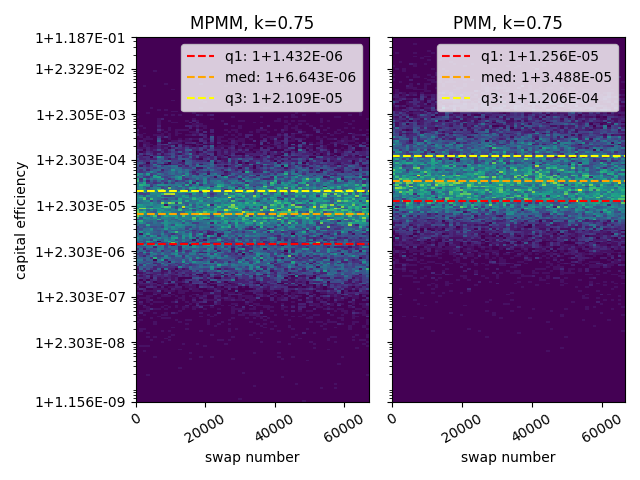}%
        \label{subfig:pmm_075_cap_eff}%
    }
    \caption{\textbf{Capital efficiency of AMMs over course of bull market simulation:} We chart the capital efficiency of swaps occurring at more than the fair market rate.}
    \label{fig:bull_cap_eff}
\end{figure*}

\begin{figure*}[t!]
    \subfloat[CPMM Variants]{%
        \includegraphics[width=.33\linewidth]{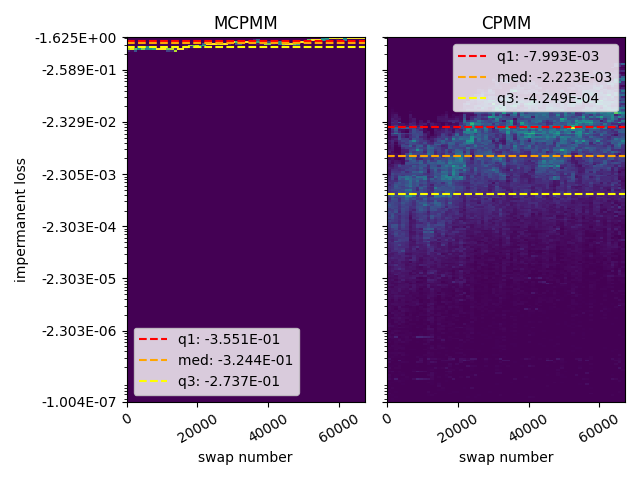}%
        \label{subfig:cpmm_imp_loss}%
    }\hfill
    \subfloat[PMM, k=0.05 variants]{%
        \includegraphics[width=.33\linewidth]{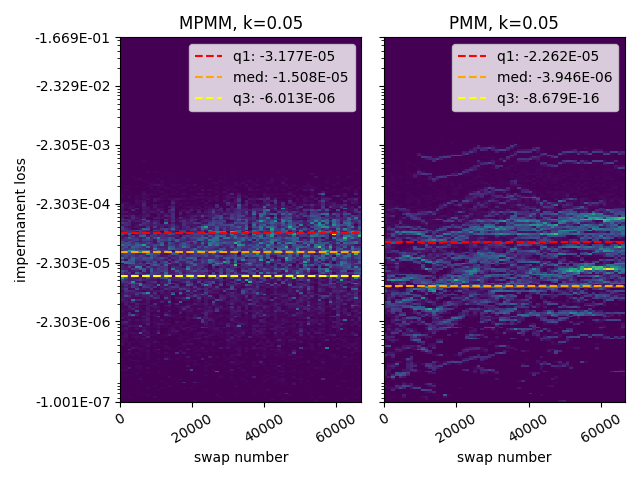}%
        \label{subfig:pmm_005_imp_loss}%
    }\hfill
    \subfloat[PMM, k=0.25 variants]{%
        \includegraphics[width=.33\linewidth]{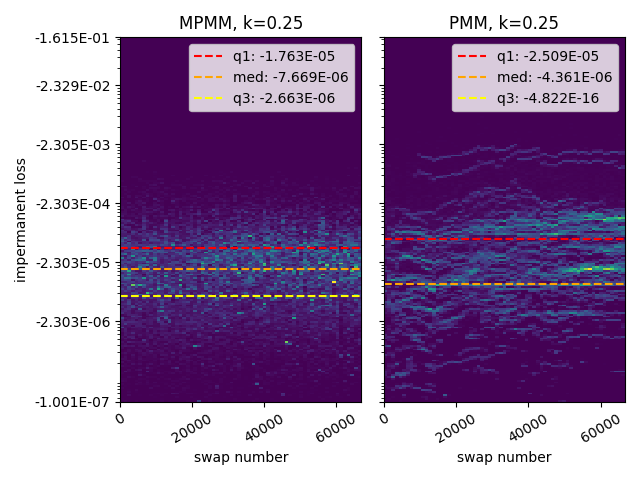}%
        \label{subfig:pmm_025_imp_loss}%
    }\\
    \subfloat[CSMM Variants]{%
        \includegraphics[width=.33\linewidth]{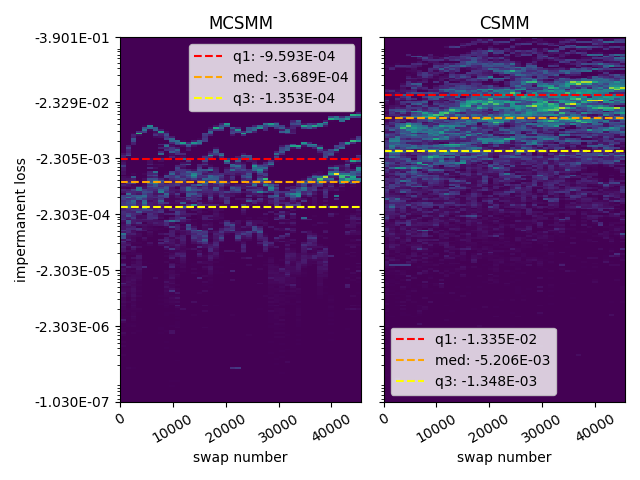}%
        \label{subfig:csmm_imp_loss}%
    }\hfill
    \subfloat[PMM, k=0.5 variants]{%
        \includegraphics[width=.33\linewidth]{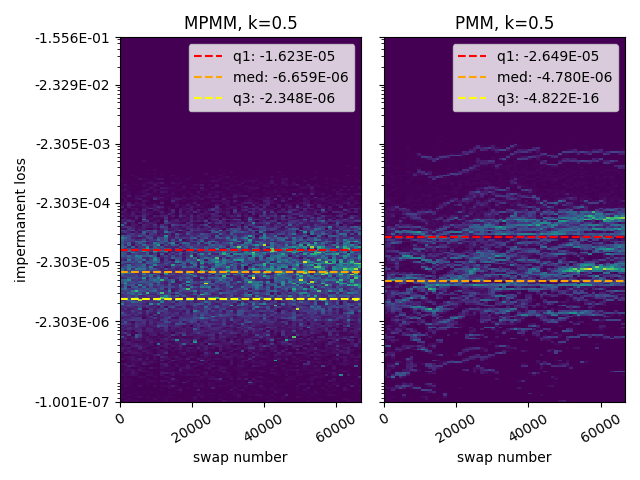}%
        \label{subfig:pmm_050_imp_loss}%
    }\hfill
    \subfloat[PMM, k=0.75 variants]{%
        \includegraphics[width=.33\linewidth]{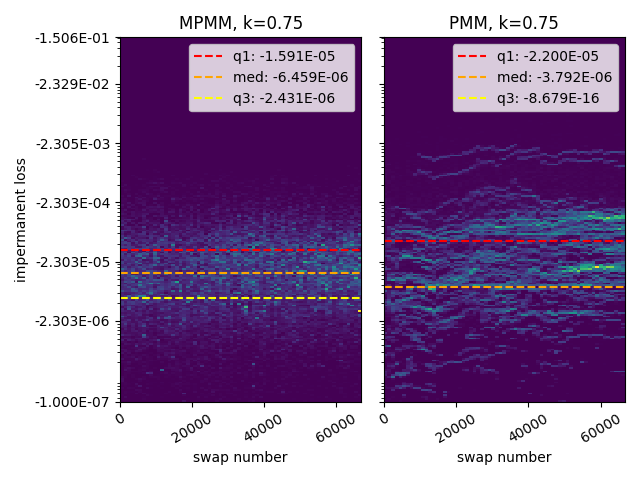}%
        \label{subfig:pmm_075_imp_loss}%
    }
    \caption{\textbf{Impermanent loss of AMMs over course of bull market simulation:} We chart the impermanent loss of pools following swaps when a loss has occurred.}
    \label{fig:bull_imp_loss}
\end{figure*}

To simulate a bull market, we perform the simulation procedure as in the previous section but only between the dates of June 22nd, 2021 and November 8th, 2021. During this period, the market cap of all crypto currencies grew 133\%, and transaction volumes grew 78\%. This scenario is an interesting challenge for AMMs in terms of impermanent loss, who should ideally match the high returns possible with a holding strategy in a bull market with significant price fluctuations. In \hyperref[fig:bull_imp_loss]{4}, we plot the simulation's impermanent losses. Now, MCPMM and MPMM perform worse than CPMM and PMM, respectively. PMM's impermanent loss significantly improves (e.g. $-3.946\mathrm{e}{-6}$ vs $-1.328\mathrm{e}{-5}$ when $k=0.05$ and $-3.792\mathrm{e}{-6}$ vs $-1.450\mathrm{e}{-5}$ when $k=0.75$ for the ``bull market" vs the ``real world price" simulation). One explanation may be that multi-token pools are slower adapting towards price changes compared to 2-token pools. CPMM style AMMs must implicitly readjust their unique equilibrium token balance (see \hyperref[sec:2.1.2]{2.1.2}) under evolving market exchange rates, and increased liquidity necessitates moving more tokens to achieve it (see \hyperref[sec:2.2]{2.2}). We hypothesize a similar principle may be hurting MPMM. Another observation is the difference between MPMM and PMM diminishes with increasing $k$: at $k=0.05$, PMM has $3.82$x fewer impermanent loss but only $1.70$x at $k=0.75$ (when comparing medians). Finally, PMM's impermanent loss varies more than MPMM (in fact, the 3rd quartile is never visible in the figure), though PMM generally has greater deviation in impermanent loss than MPMM across all of this paper's simulations. We believe this is simply because 2-token AMMs were initialized with significantly more pools than multi-token AMMs. Finally, in \hyperref[fig:bull_cap_eff]{3}, we still see similar capital efficiency trends as the prior simulation from the previous section.

\subsection{Bear Market Simulation}
\begin{figure*}[t!]
    \subfloat[CPMM variants]{%
        \includegraphics[width=.33\linewidth]{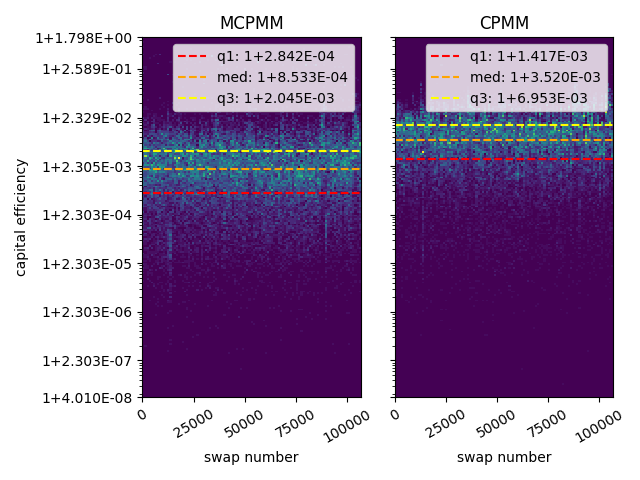}%
        \label{subfig:cpmm_cap_eff}%
    }\hfill
    \subfloat[PMM, k=0.05 variants]{%
        \includegraphics[width=.33\linewidth]{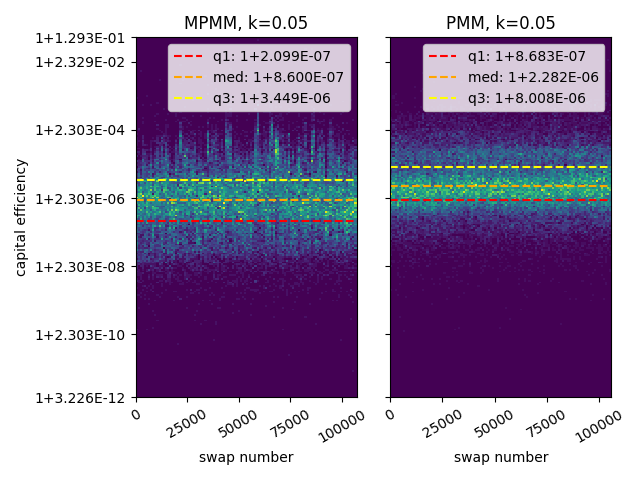}%
        \label{subfig:pmm_005_cap_eff}%
    }\hfill
    \subfloat[PMM, k=0.25 variants]{%
        \includegraphics[width=.33\linewidth]{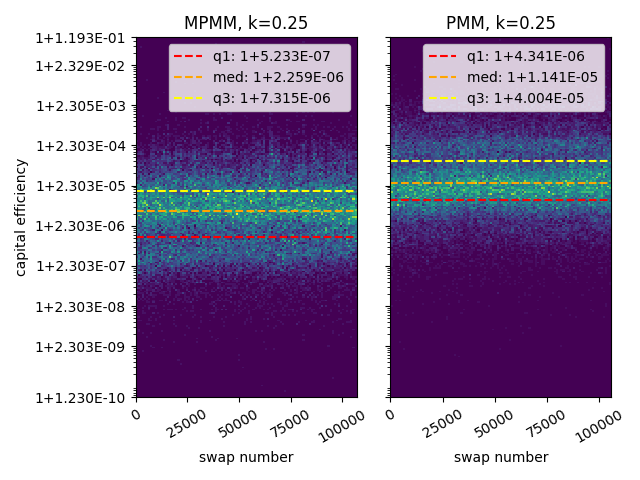}%
        \label{subfig:pmm_025_cap_eff}%
    }\\
    \subfloat[CSMM variants]{%
        \includegraphics[width=.33\linewidth]{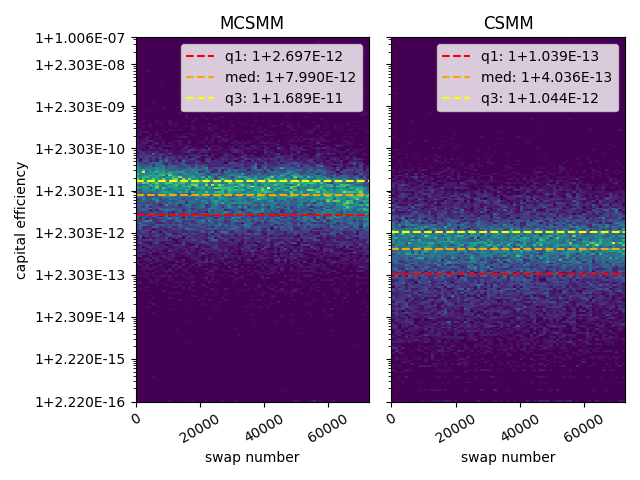}%
        \label{subfig:csmm_cap_eff}%
    }\hfill
    \subfloat[PMM, k=0.5 variants]{%
        \includegraphics[width=.33\linewidth]{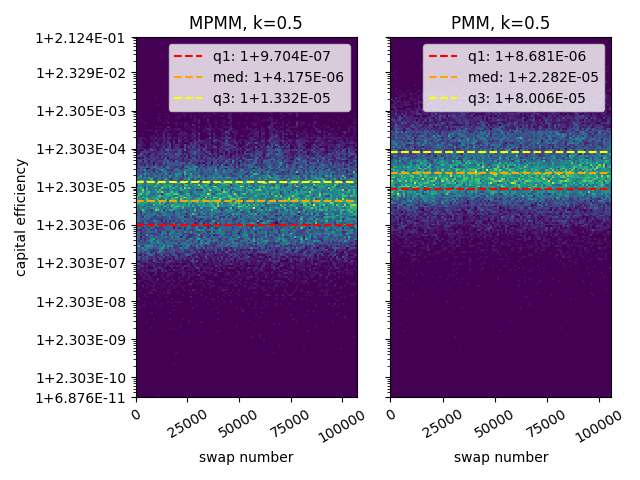}%
        \label{subfig:pmm_050_cap_eff}%
    }\hfill
    \subfloat[PMM, k=0.75 variants]{%
        \includegraphics[width=.33\linewidth]{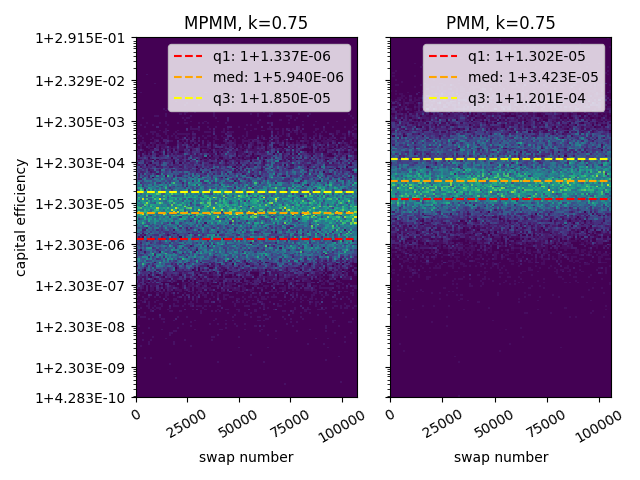}%
        \label{subfig:pmm_075_cap_eff}%
    }
    \caption{\textbf{Capital efficiency of AMMs over course of bear market simulation:} We chart the capital efficiency of swaps occurring at more than the fair market rate.}
    \label{fig:bear_cap_eff}
\end{figure*}

\begin{figure*}[t!]
    \subfloat[CPMM Variants]{%
        \includegraphics[width=.33\linewidth]{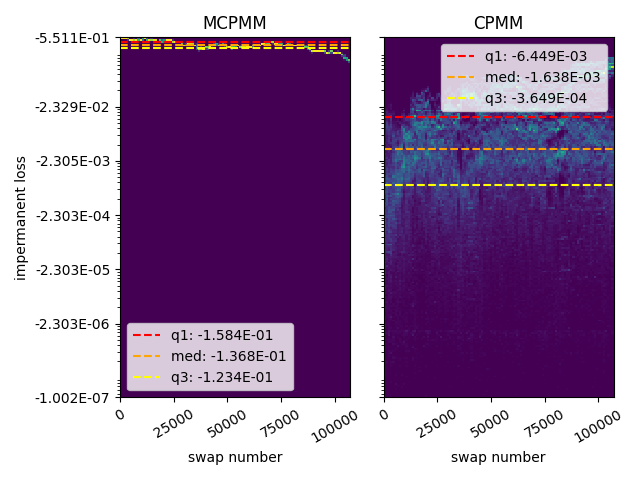}%
        \label{subfig:cpmm_imp_loss}%
    }\hfill
    \subfloat[PMM, k=0.05 variants]{%
        \includegraphics[width=.33\linewidth]{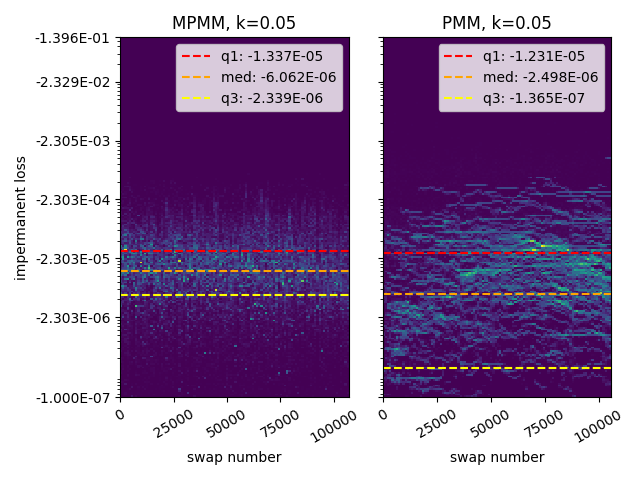}%
        \label{subfig:pmm_005_imp_loss}%
    }\hfill
    \subfloat[PMM, k=0.25 variants]{%
        \includegraphics[width=.33\linewidth]{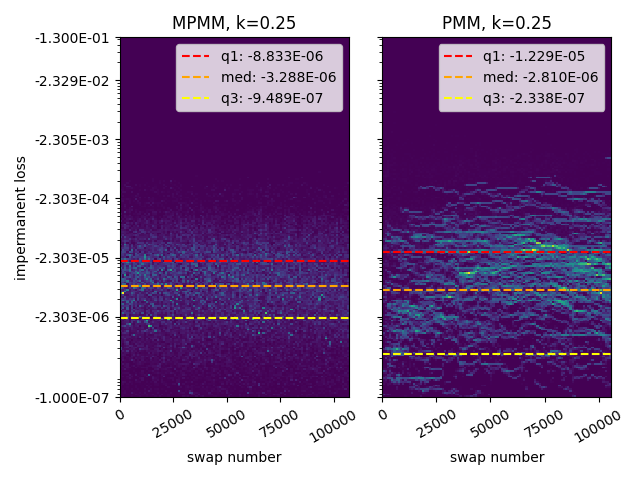}%
        \label{subfig:pmm_025_imp_loss}%
    }\\
    \subfloat[CSMM Variants]{%
        \includegraphics[width=.33\linewidth]{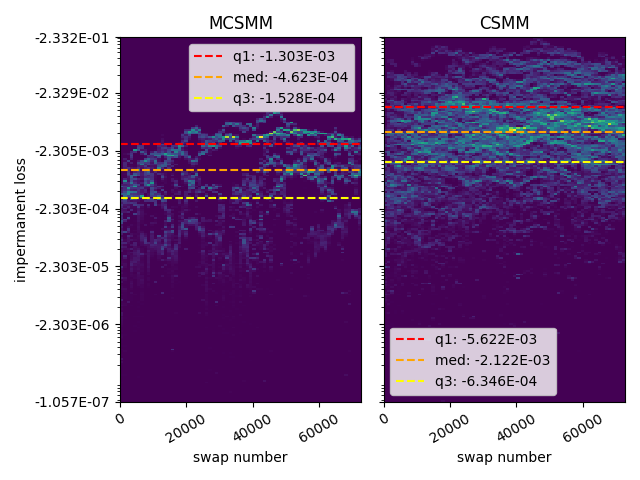}%
        \label{subfig:csmm_imp_loss}%
    }\hfill
    \subfloat[PMM, k=0.5 variants]{%
        \includegraphics[width=.33\linewidth]{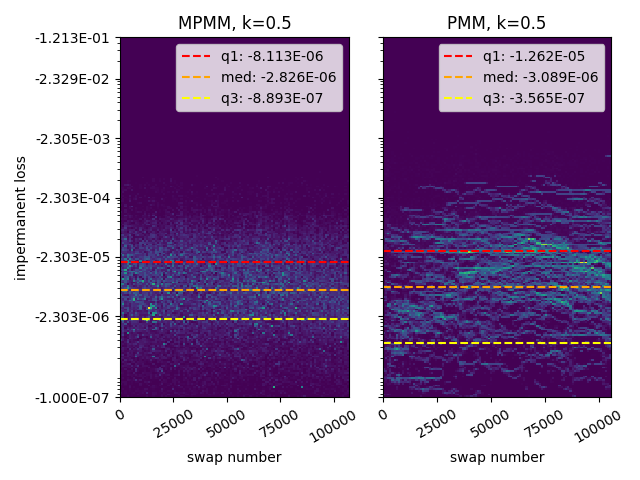}%
        \label{subfig:pmm_050_imp_loss}%
    }\hfill
    \subfloat[PMM, k=0.75 variants]{%
        \includegraphics[width=.33\linewidth]{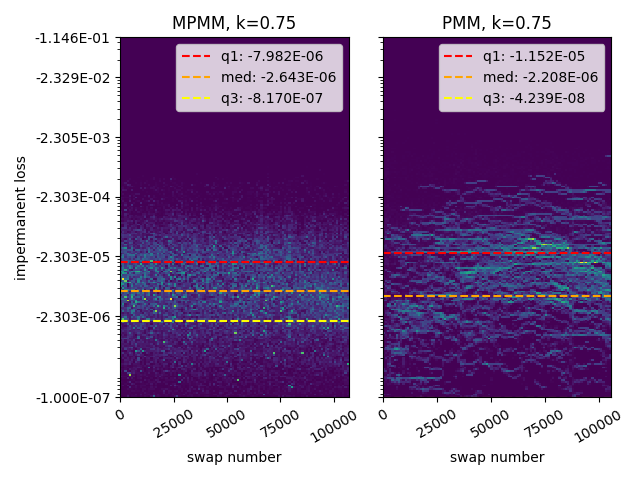}%
        \label{subfig:pmm_075_imp_loss}%
    }
    \caption{\textbf{Impermanent loss of AMMs over course of bear market simulation:} We chart the impermanent loss of pools following swaps when a loss has occurred.}
    \label{fig:bear_imp_loss}
\end{figure*}

The bear market is simulated similarly to prior simulations. The dates now range between November 8th, 2021 to June 18th, 2022. During this period, trading volume fell by 67\% and the total crypto market cap by 64\%. This represents the most ideal scenario for impermanent loss, where AMMs need to beat a heavily discounted holding portfolio. From \hyperref[fig:bull_imp_loss]{6}, that is exactly what is observed: all AMMs experience lower impermanent loss than in the extended period simulation \hyperref[fig:normal_imp_loss]{2} and in the bull market \hyperref[fig:bull_imp_loss]{4}. Also, while PMM still typically has lower median impermanent loss than MPMM, this gap is significantly closer than in the bull market simulation. Finally, \hyperref[fig:bull_cap_eff]{5} shows capital efficiency plotted similarly to previous sections.

\subsection{Token Crash Scenario}
\begin{figure*}[t!]
    \subfloat[CPMM variants]{%
        \includegraphics[width=.33\linewidth]{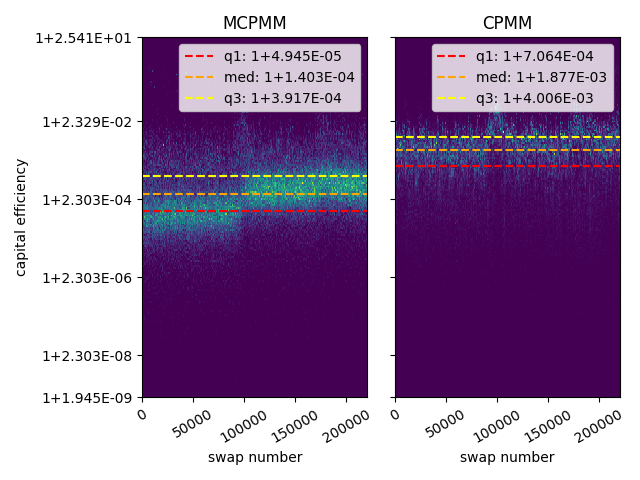}%
        \label{subfig:cpmm_cap_eff}%
    }\hfill
    \subfloat[PMM, k=0.05 variants]{%
        \includegraphics[width=.33\linewidth]{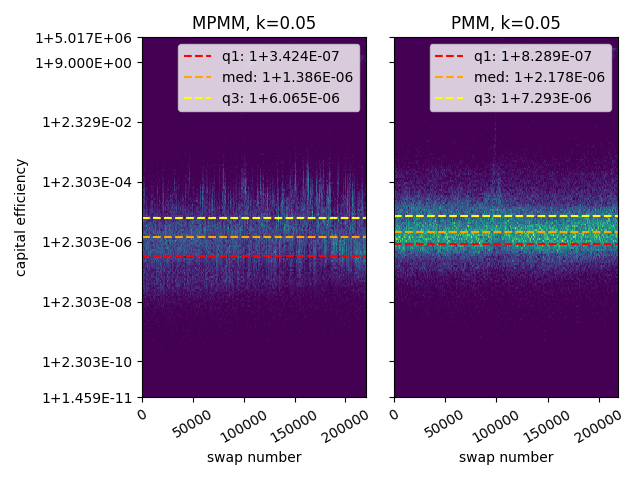}%
        \label{subfig:pmm_005_cap_eff}%
    }\hfill
    \subfloat[PMM, k=0.25 variants]{%
        \includegraphics[width=.33\linewidth]{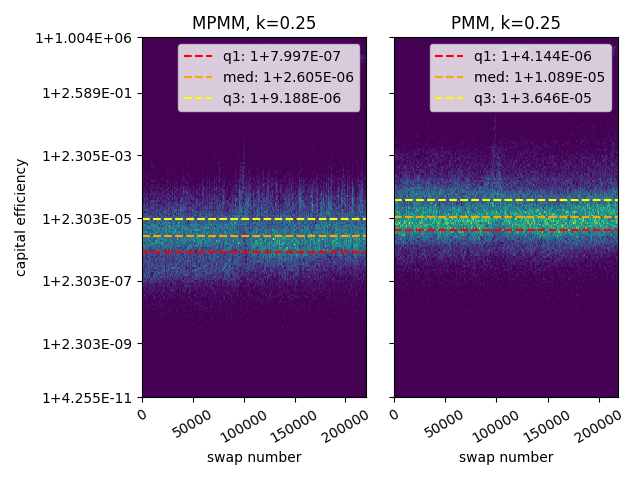}%
        \label{subfig:pmm_025_cap_eff}%
    }\\
    \subfloat[CSMM variants]{%
        \includegraphics[width=.33\linewidth]{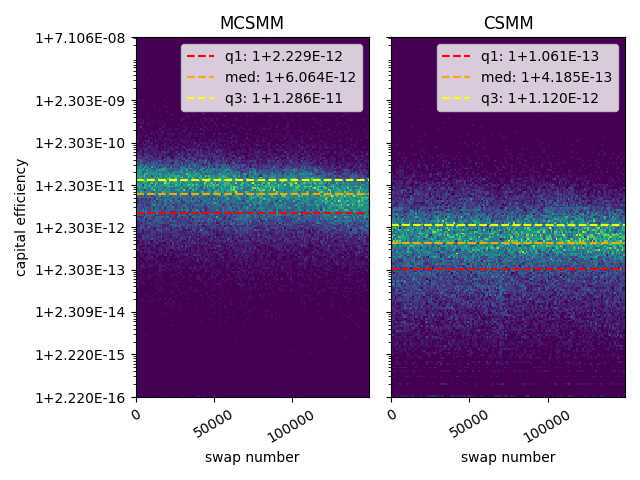}%
        \label{subfig:csmm_cap_eff}%
    }\hfill
    \subfloat[PMM, k=0.5 variants]{%
        \includegraphics[width=.33\linewidth]{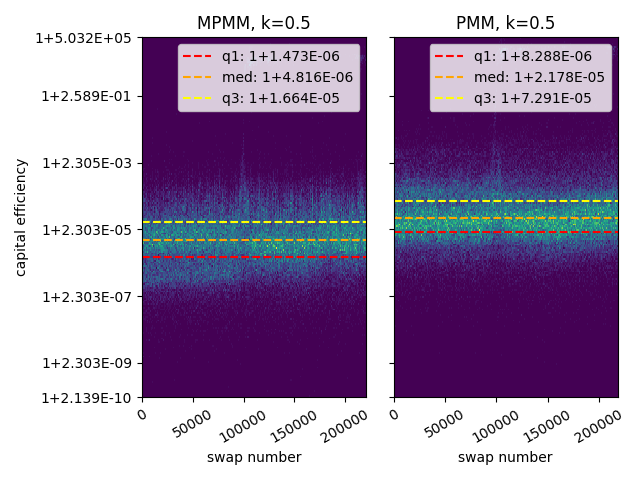}%
        \label{subfig:pmm_050_cap_eff}%
    }\hfill
    \subfloat[PMM, k=0.75 variants]{%
        \includegraphics[width=.33\linewidth]{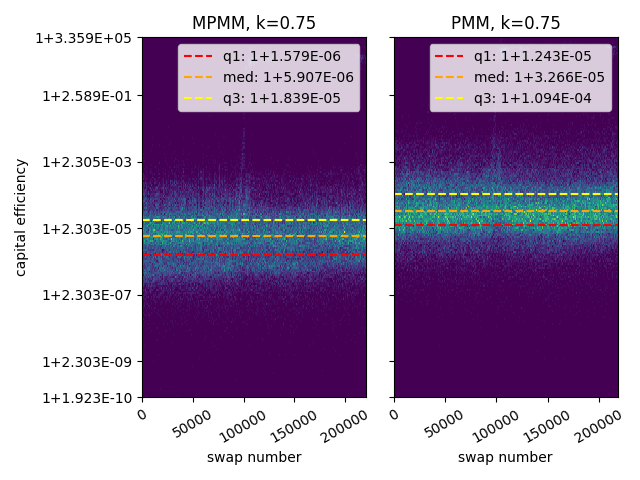}%
        \label{subfig:pmm_075_cap_eff}%
    }
    \caption{\textbf{Capital efficiency of AMMs over course of a token crash simulation:} We chart the capital efficiency of swaps occurring at more than the fair market rate.}
    \label{fig:crash_cap_eff}
\end{figure*}

\begin{figure*}[t!]
    \subfloat[CPMM Variants]{%
        \includegraphics[width=.33\linewidth]{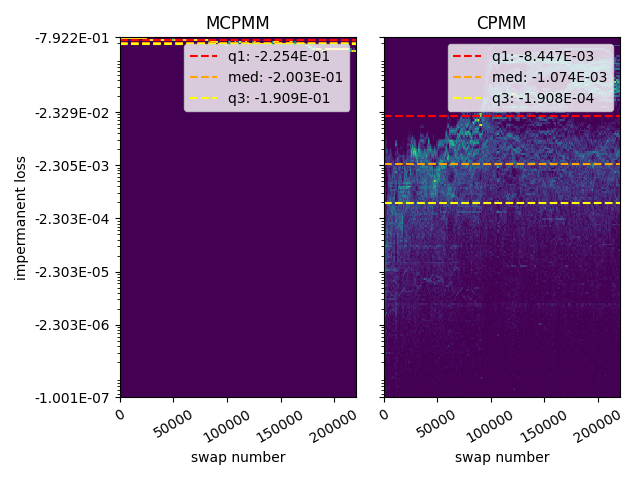}%
        \label{subfig:cpmm_imp_loss}%
    }\hfill
    \subfloat[PMM, k=0.05 variants]{%
        \includegraphics[width=.33\linewidth]{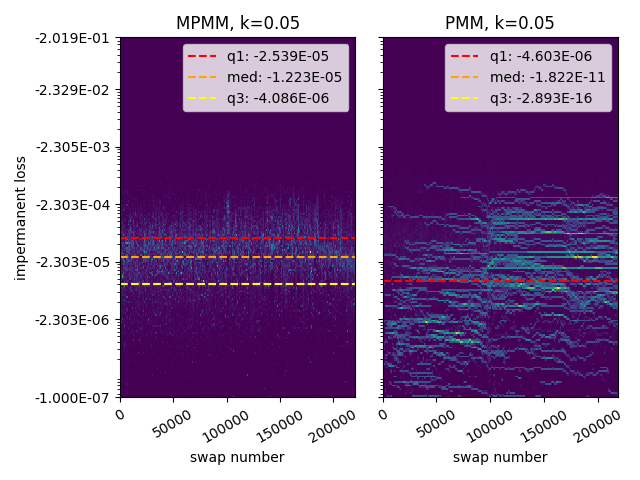}%
        \label{subfig:pmm_005_imp_loss}%
    }\hfill
    \subfloat[PMM, k=0.25 variants]{%
        \includegraphics[width=.33\linewidth]{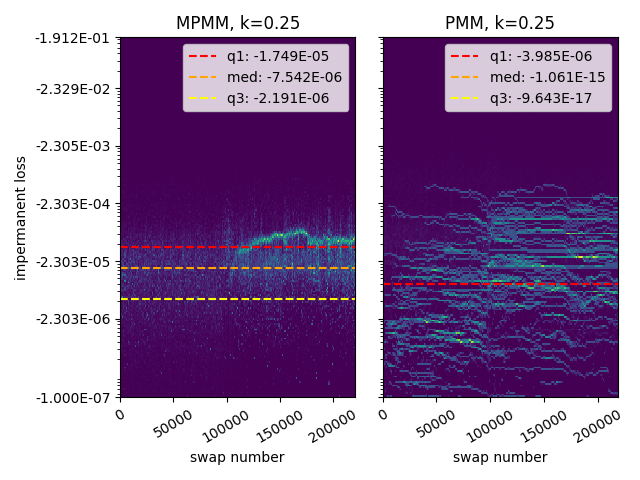}%
        \label{subfig:pmm_025_imp_loss}%
    }\\
    \subfloat[CSMM Variants]{%
        \includegraphics[width=.33\linewidth]{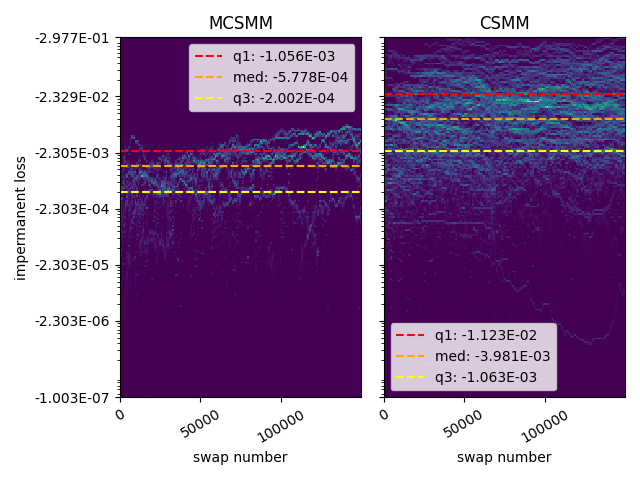}%
        \label{subfig:csmm_imp_loss}%
    }\hfill
    \subfloat[PMM, k=0.5 variants]{%
        \includegraphics[width=.33\linewidth]{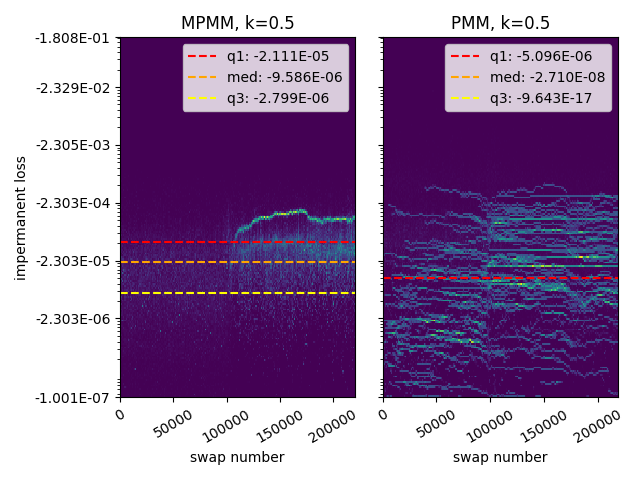}%
        \label{subfig:pmm_050_imp_loss}%
    }\hfill
    \subfloat[PMM, k=0.75 variants]{%
        \includegraphics[width=.33\linewidth]{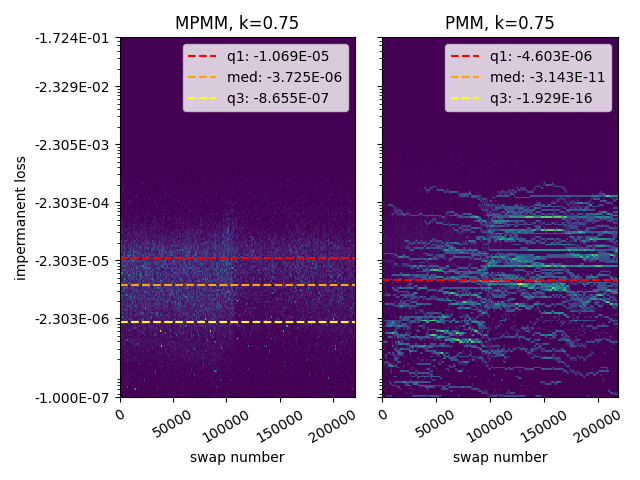}%
        \label{subfig:pmm_075_imp_loss}%
    }
    \caption{\textbf{Impermanent loss of AMMs over course of a token crash simulation:} We chart the impermanent loss of pools following swaps when a loss has occurred.}
    \label{fig:crash_imp_loss}
\end{figure*}

Finally, we study a scenario where tokens' prices drastically drop. We perform a simulation identical to first scenario, but also add Terra Luna Classic and USTC (formerly UST) to the simulation (we did not have hourly pricing data for either token, but estimate it similarly to the hourly volume data). We run the simulation from April 1st, 2022 to July 1st, 2022 to capture the period immediately before and after the tokens' crashes. Additionally, we change the number of hourly swaps to 100 to increase the number of Luna and USTC swaps during the crash. This last scenario tests the susceptibility of multi-token AMMs to extreme volatility in a few tokens, which 2-token pool AMMs explicitly silo to only affect pools containing the volatile token. In \hyperref[fig:crash_imp_loss]{8}, we see in some AMMs an uptick in impermanent loss around the 100,000th swap, which corresponds to early May 2022, shortly after the worst of the crash. The effect is more pronounced in 2-token pool AMMs (likely the pools containing the failing tokens), which confirms that smaller liquidity pools are more susceptible to impermanent loss during volatile markets. However, when considering the median impermanent loss across all pools, multi-token AMMs suffer significantly compared to 2-token counterparts, showing that price shocks easily affect multi-token pools while 2-token pools isolate volatility to only the trading pairs containing the failing token. Finally, when plotting the capital efficiency in  \hyperref[fig:crash_cap_eff]{7}, we also observe a spike in inefficiency surrounding the 100,000th swap. One explanation might be that token prices in simulation during the crash decrease slower than in our source data: in the real world, prices seem to have dropped exponentially over time while we only linearly interpolate prices between consecutive days to obtain hourly prices.

\section{Conclusion and Further Work}

In this paper, we propose the Multi-Token Proactive Market Maker (MPMM), which adopts principles from DODO Exchange's PMM. We analyze PMM's use of a parameter $k$ to ``interpolate" an exchange rate curve between the naive constant sum (CSMM) and constant product (CPMM) market makers. Similar to CPMM, PMM pools are not drainable and have equilibrium points matching LPs' deposit amounts. Similar to CSMM, PMM pools can match fair market exchange rates throughout much of its liquidity range, thus preventing exploits surrounding exchange rate movements such as arbitrage and front-running. PMM also introduces a recovery operation to reduce impermanent losses during exchange rate changes. Through our analysis of PMM, we show for PMM that 1. \textit{increasing liquidity helps maintain exchange rates near the market rate across a larger liquidity spectrum} and 2. \textit{makes the impermanent loss recovery procedure more robust}, leading to the conception of multi-token pools that combine 2-token pools in traditional AMMs to access more liquidity. We then focus on constructing MPMM, the multi-token generalization of PMM, via a simple objective of maintaining LPs deposits while allowing arbitrary swapping between token types. Through simulation under a variety of market scenarios, we verify that MPMM has better capital efficiency and impermanent loss performance than other 2 and multi-token AMMs.
\\\\
One area that could be further investigated is reducing the propagation of losses across a multi-token pool. An easy solution would be to forbid swapping when token prices have changed drastically within a period of time. However, it may be challenging to define a threshold of volatility that protects liquidity providers' funds while maintaining ease of trading. Another alternative may be to incorporate price prediction mechanisms. For example, the ARIMA model was successfully used to predict Bitcoin closing prices~\cite{price_pred}. However, similar methods may take significant testing to calibrate or are too computationally expensive (for example, requiring tracking a long log of historical prices or metrics).
\\\\
Another concept to explore is generalizing the PMM balance curve to a convex piecewise-defined multi-dimensional surface for multi-token pools. Such a surface should be nearly flat near some equilibrium balance point and should asymptotically approach all coordinate planes whenever any token balance is excessively large. Additionally, it is desirable for the surface to be parameterized by some constant number of variables to ensure ease of computations on a blockchain.

\section*{Contributions}
This work was mainly led by Wayne Chen. He was responsible for the project's conceptualization, methodology, software, formal analysis, writing, visualization, and administration. Songwei Chen contributed to the conceptualization and administration. Finally, Preston Rozwood assisted in conceptualization, methodology, writing, and supervision.

\section*{Acknowledgements}
We'd also like to thank Stanley Jiang, Krishna Palempalli, Michael Mirkin, Fan Zhang, Bineet Mishra, Weizhao Tang, Haoqian Zhang, and Marwa Mouallem for writing parts of the simulation code. Additionally, we want to thank the \href{https://www.initc3.org/}{Initiative for Cryptocurrencies and Contracts} for allowing us to present our initial findings.

\section*{Declaration of Competing Interests}

The authors have no competing interests that affect this work.

\section*{Sources of Funding}

This research did not receive any specific grant from funding agencies in the public, commercial, or not-for-profit sectors. 

\bibliographystyle{plain}
\bibliography{citations}

\pagebreak

\section{Appendix}
\subsection{Proof of Equation 1}\label{sec:proof_theorem_1}

First, we prove that only $0 < B < B_0$ or $0 < Q \leq Q_0$, or $B = B_0$ and $Q = Q_0$ simultaneously.
The last case is trivial after substituting $B = B_0$ into \hyperref[eq:system_1]{1a} and $Q = Q_0$ into \hyperref[eq:system_1]{1b}. Suppose for a contradiction that both $0 < B < B_0$ and $0 < Q \leq Q_0$. It follows from \hyperref[eq:system_1]{1a} that $Q > Q_0$. Similarly, assuming that $0 < Q < Q_0$ results in $B > B_0$.
\\
\\
We now find $\frac{\partial Q}{\partial B}$ and $\frac{\partial B}{\partial Q}$ for when $0 < B \leq B_0$. The derivations for when $0 < Q \leq Q_0$ follow similarly.

\begin{align*}
    \frac{\partial Q}{\partial B} &= \frac{\partial}{\partial B}\left[-\frac{P_B}{P_Q}(B - B_0)\left(1 - k + \frac{kB_0}{B} \right) + Q_0 \right]\\
    &= -\frac{P_B}{P_Q}\left[(B-B_0) \cdot -\frac{kB_0}{B^2} + \left(1 - k + \frac{kB_0}{B} \right) \right]\\
    &= \frac{P_B}{P_Q} \cdot \frac{kB_0}{B^2}\left(-\frac{B^2}{kB_0} + \frac{B^2}{B_0} - B_0 \right)\\
    &= \frac{P_B}{P_Q} \cdot \frac{1}{B^2}(-B^2 + kB^2 -kB_0^2)\\
    &= \frac{P_B(B_0^2k - (k-1)B^2)}{P_Q B^2}
\end{align*}

To find $\frac{\partial B}{\partial Q}$, we write \hyperref[eq:system_1]{1a} as a function of $Q$.

\begin{align*}
    Q &= -\frac{P_B}{P_Q}(B - B_0)\left(1 - k + \frac{kB_0}{B} \right) + Q_0\\
    &= -\frac{P_B}{P_Q}(B - B_0)\left(\frac{-kB + kB_0 + B}{B} \right) + Q_0\\
    &= \frac{-P_B(B-B_0)(-kB + kB_0 + B) + Q_0 P_Q B}{P_Q B}\\
    \\
    0 &= P_B B^2k - 2P_B B k B_0 - P_B B^2 + P_B B_0^2 k + P_B B_0 B + Q_0 P_Q B - P_Q B Q\\
    &= B^2(P_B k - P_B) + B(P_B B_0 - 2P_B B_0 k + Q_0 P_Q - P_Q Q) + P_B B_0^2 k\\
    &= B^2(P_B(k-1)) + B(P_B B_0 (1 - 2k) + P_Q (Q_0 - Q)) + P_B B_0^2 k\\
    \\
    B &= \frac{P_B B_0 (2k - 1) + P_Q (Q - Q_0) \pm \sqrt{(P_B B_0 (1 - 2k) + P_Q (Q_0 - Q))^2 - 4 P_B^2 B_0^2 k (k-1)}}{2P_B(k-1)}
\end{align*}

We know that $B = B_0$ and $Q = Q_0$ is a solution to this equation. We verify the correct root via checking

\begin{align*}
    B_0 &= \frac{P_B B_0 (2k - 1) + P_Q (Q_0 - Q_0) \pm \sqrt{(P_B B_0 (1 - 2k) + P_Q (Q_0 - Q_0))^2 - 4 P_B^2 B_0^2 k (k-1)}}{2P_B(k-1)}\\
    &= \frac{P_B B_0 (2k - 1) \pm \sqrt{P_B^2 B_0^2 (1 - 2k)^2 - 4 P_B^2 B_0^2 k (k-1)}}{2P_B(k-1)}\\
    &= \frac{P_B B_0 \left(2k - 1 \pm \sqrt{1 - 4k + 4k^2 - 4k^2 + 4k} \right)}{P_B(2k-2)}\\
    &= \frac{B_0(2k - 1 \pm 1)}{2k-2}
\end{align*}

which only holds for the negative root. From here, we take derivatives:

\begin{align*}
    \frac{\partial B}{\partial Q} &= \frac{\partial}{\partial Q}\left[\frac{P_B B_0 (2k - 1) + P_Q (Q - Q_0) - \sqrt{(P_B B_0 (1 - 2k) + P_Q (Q_0 - Q))^2 - 4 P_B^2 B_0^2 k (k-1)}}{2P_B(k-1)} \right]\\
    &= \frac{1}{2P_B(k-1)}(P_Q - \frac{2(P_B B_0 (1 - 2k) + P_Q (Q_0 - Q)) \cdot -P_Q}{2\sqrt{(P_B B_0 (1 - 2k) + P_Q (Q_0 - Q))^2 - 4 P_B^2 B_0^2 k (k-1)}})\\
    &= \frac{P_Q + \frac{P_Q(P_B B_0 (1 - 2k) + P_Q (Q_0 - Q))}{\sqrt{(P_B B_0 (1 - 2k) + P_Q (Q_0 - Q))^2 - 4 P_B^2 B_0^2 k (k-1)}}}{2P_B(k-1)}
\end{align*}

\subsection{Proof of 2.1.1}\label{sec:proof:2_1_1}

We prove the first case only, and the second follows similarly.

\begin{align*}
  \lim_{\frac{P_Q}{P_B} \rightarrow \infty} \frac{\partial Q}{\partial B} &= \lim_{\frac{P_Q}{P_B} \rightarrow \infty} \left(P_B + \frac{P_B(P_Q Q_0 (1 - 2k) + P_B (B_0 - B))}{\sqrt{(P_Q Q_0 (1 - 2k) + P_B (B_0 - B))^2 - 4 P_Q^2 Q_0^2 k (k-1)}} \right) \cdot (2P_Q(k-1))^{-1}\\
  &= \lim_{\frac{P_Q}{P_B} \rightarrow \infty} \frac{P_B}{P_Q}\left(1 + \frac{P_Q Q_0 (1 - 2k) + P_B (B_0 - B)}{\sqrt{(P_Q Q_0 (1 - 2k) + P_B (B_0 - B))^2 - 4 P_Q^2 Q_0^2 k (k-1)}} \right)\\
  &= \lim_{\frac{P_Q}{P_B} \rightarrow \infty} \frac{P_B}{P_Q}\left(1 + \frac{Q_0 (1 - 2k) + \frac{P_B}{P_Q} (B_0 - B)}{\sqrt{(Q_0 (1 - 2k) + \frac{P_B}{P_Q} (B_0 - B))^2 - 4 Q_0^2 k (k-1)}} \right)\\
  &= \lim_{\frac{P_Q}{P_B} \rightarrow \infty} \frac{P_B}{P_Q}\left(1 + \frac{Q_0 (1 - 2k)}{\sqrt{(Q_0 (1 - 2k))^2 - 4 Q_0^2 k (k-1)}} \right)\\
  &= \lim_{\frac{P_Q}{P_B} \rightarrow \infty} \frac{P_B}{P_Q}\left(1 + \frac{1 - 2k}{\sqrt{1 - 4k + 4k^2 - 4k^2 + 4k}} \right)\\
  &= \lim_{\frac{P_Q}{P_B} \rightarrow \infty} \frac{P_B}{P_Q}(2 - 2k)\\
  &= 0
\end{align*}

\subsection{Proof of 2.1.2}\label{sec:proof:2_1_2}

We prove that $\lim_{Q \rightarrow Q_0} \frac{\partial B}{\partial Q} = \frac{P_Q}{P_B}$ and $\frac{\partial B}{\partial Q} > \frac{P_Q}{P_B}$ for when $Q < Q_0$, and the analogous results for $B < B_0$ follow similarly. From \hyperref[sec:proof_theorem_1]{8.1}, we know that $Q < Q_0$ and $B \leq B_0$ are the only cases to consider.

\begin{align*}
    \lim_{Q \rightarrow Q_0} \frac{\partial B}{\partial Q} &= \lim_{Q \rightarrow Q_0} \frac{P_Q(Q_0^2k - (k-1)Q^2)}{P_B Q^2}\\
    &= \lim_{Q \rightarrow Q_0} \frac{P_Q}{P_B} \cdot \frac{Q_0^2k - (k-1)Q^2}{Q^2}\\
    &= \lim_{Q \rightarrow Q_0} \frac{P_Q}{P_B} \cdot \left(k\left(\left(\frac{Q_0}{Q} \right)^2 - 1 \right) + 1 \right)\\
    &= \frac{P_Q}{P_B}
\end{align*}

\begin{align*}
    \frac{\partial B}{\partial Q} &= \frac{P_Q}{P_B} \cdot \left(k\left(\left(\frac{Q_0}{Q} \right)^2 - 1 \right) + 1 \right)\\
    &> \frac{P_Q}{P_B} \cdot \left(k\left(\left(\frac{Q_0}{Q_0} \right)^2 - 1  \right) + 1 \right)\\
    &= \frac{P_Q}{P_B}
\end{align*}

\subsection{More Notes Regarding 2.1.3}\label{sec:proof:2_1_3}

DODO's invariant for calculating the new equilibrium points is that the token in excess (with respect to the original equilibrium point) does not decrease its equilibrium point. Therefore, the excess token's new equilibrium is identical to its original while the depleted token's new equilibrium is determined by this and all other constraints of the PMM curve formulation. The overall effect is deriving a new PMM curve that passes through the current balance point while maximizing the amount of the depleted token that can be recouped with only the excess amount (with respect to the curret equilibrium) of the excess token~\cite{dodo_docs}.
\\
\\
We consider only the case in \hyperref[eq:equilibrium_short]{4}, since  \hyperref[eq:equilibrium_long]{5} follows similarly. Since $B < B_0$, $Q_{0'} = Q_0$, and the token prices are now $P_{B'}$ and $P_{Q'}$ we solve the following for $B_{0'}$:

\begin{align*}
    Q &= -\frac{P_{B'}}{P_{Q'}}(B - B_{0'})\left(1 - k + \frac{kB_{0'}}{B} \right) + Q_0\\
    &= -\frac{P_{B'}}{P_{Q'}}(B - B_{0'})\left(\frac{kB_{0'} + B(1-k)}{B} \right) + Q_0\\
    &= \frac{-P_{B'}(B-B_{0'})(kB_{0'} + B(1-k)) + Q_0 P_{Q'} B}{P_{Q'} B}\\
    &= \frac{-P_{B'}(-B_{0'}^2 k + B_{0'}(kB - B(1-k)) + B^2(1-k)) + Q_0 P_{Q'} B}{P_{Q'} B}\\
    &= \frac{P_{B'}k}{P_{Q'}B} B_{0'}^2 - \frac{P_{B'}B(2k-1)}{P_{Q'}B} B_{0'} + \frac{Q_0 P_{Q'} B + P_{B'} B^2 (k-1)}{P_{Q'}B}\\
    \\
    B_{0'} &= \left(\frac{P_{B'}B(2k-1)}{P_{Q'}B} \pm \sqrt{\left(\frac{P_{B'}B(2k-1)}{P_{Q'}B} \right)^2 - 4 \cdot \frac{P_{B'}k}{P_{Q'}B} \cdot \frac{Q_0 P_{Q'} B + P_{B'} B^2 (k-1) - Q P_{Q'} B}{P_{Q'}B}} \right) \cdot \left(2 \cdot \frac{P_{B'}k}{P_{Q'}B} \right)^{-1}\\
    &= \frac{2P_{B'}Bk - P_{B'}B \pm \sqrt{P_{B'}^2 B^2 (2k-1)^2 - 4 P_{B'}k (Q_0 P_{Q'} B + P_{B'} B^2 (k-1) - Q P_{Q'} B)}}{2 P_{B'}k}\\
    &= B - \frac{B}{2k} \pm \sqrt{\frac{P_{B'}^2 B^2 ((2k-1)^2 - 4k(k-1)) - 4 P_{B'} P_{Q'} B k (Q_0 - Q)}{4 P_{B'}^2 k^2}}\\
    &= B - \frac{B}{2k} \pm \sqrt{\frac{B^2 (4k^2 - 4k + 1 - 4k^2 + 4k)}{4 k^2} - \frac{P_{Q'} B (Q_0 - Q)}{P_{B'} k}}\\
    &= B - \frac{B}{2k} \pm \sqrt{\frac{B^2}{4 k^2} - \frac{4k}{B} \cdot \frac{B^2 (Q_0 - Q)}{4 \frac{P_{B'}}{P_{Q'}} k^2}}\\
    &= B - \frac{B}{2k} \pm \sqrt{\frac{B^2}{4k^2}\left(1 - \frac{4 k (Q_0 - Q)}{\frac{P_{B'}}{P_{Q'}} B} \right)}\\
    &= B + \frac{B}{2k} \left(\pm \sqrt{1 + \frac{4k(Q - Q_0)}{\frac{P_{B'}}{P_{Q'}} B}} - 1 \right)
\end{align*}

We cannot use \hyperref[eq:system_1]{1a} to verify the correct root, but since the negative root yields a negative $B_{0'}$ for sufficiently large $Q$, the positive root is the solution.

\subsection{CPMM and CSMM in 3.1.4}\label{sec:proof_3_1_4}

The CSMM formula requires that a weighted sum of the token balances be always equal to a constant. \hyperref[eq:csmm]{6} suggests that the constant and weights are dependent on price and an equilibrium token balances, but these could be fixed to the initial exchange rate and starting token balances, respectively. Since the most relevant property of CSMM for capital efficiency and impermanent loss is related solely to its constant exchange rate at all token balances, and fixing these values in place would guarantee that a token be immediately arbitraged away as soon as exchange rates changed, our simulations only enforced a constant exchange rate matching the market rate and did not fix weights or the constant.
\\
\\
The CPMM formula requires the product of the token balances be always equal to a constant. We can rewrite \hyperref[eq:cpmm]{7a} as follows to make this clearer:

\begin{align*}
    Q &= -\frac{P_B}{P_Q}(B - B_0)\frac{B_0}{B} + Q_0\\
    \frac{B}{B_0} (Q - Q_0) &= -\frac{P_B}{P_Q} (B - B_0)\\
    \frac{B Q}{B_0} - \frac{B Q_0}{B_0} &= \frac{P_B B_0}{P_Q} - \frac{P_B B}{P_Q}\\
    B\left(\frac{Q}{B_0} - \frac{Q_0}{B_0} + \frac{P_B}{P_Q} \right) &= \frac{P_B B_0}{P_Q}
\end{align*}

This expression is of the form $x(ay + b) = c$, where $a$, $b$, and $c$ are constants. So, this is just a typical $xy = c$ CPMM curve that's been shifted and scaled.

\subsection{$\frac{\partial Q}{\partial B} \rightarrow \frac{P_B}{P_Q}$ as $k \rightarrow 0$ for fixed $B$ and $\frac{\partial B}{\partial Q} \rightarrow \frac{P_Q}{P_B}$ as $k \rightarrow 0$ for fixed $Q$}\label{sec:proof_small_k}
We prove the statement using the equations given in \hyperref[eq:marginals_short]{2} and the one in \hyperref[eq:marginals_long]{3} follows similarly.

\begin{align*}
    \lim_{k \rightarrow 0} \frac{\partial Q}{\partial B} &= \lim_{k \rightarrow 0} \frac{P_B(B_0^2k - (k-1)B^2)}{P_Q B^2}\\
    &= \lim_{k \rightarrow 0} \frac{P_B \cdot B^2}{P_Q \cdot B^2}\\
    &= \frac{P_B}{P_Q}
\end{align*}

\begin{align*}
    \lim_{k \rightarrow 0} \frac{\partial B}{\partial Q} &= \lim_{k \rightarrow 0} \left(P_Q + \frac{P_Q(P_B B_0 (1 - 2k) + P_Q (Q_0 - Q))}{\sqrt{(P_B B_0 (1 - 2k) + P_Q (Q_0 - Q))^2 - 4 P_B^2 B_0^2 k (k-1)}} \right) \cdot (2P_B(k-1))^{-1}\\
    &= \lim_{k \rightarrow 0} \left(P_Q + \frac{P_Q(P_B B_0 + P_Q (Q_0 - Q))}{\sqrt{(P_B B_0 + P_Q (Q_0 - Q))^2}} \right) \cdot (-2P_B)^{-1}\\
    &= \lim_{k \rightarrow 0} \frac{P_Q + P_Q}{-2P_B}\\
    &= -\frac{P_Q}{P_B}
\end{align*}

\subsection{Increased Capital Efficiency with Increased Liquidity}\label{sec:liquidity_boosts_cap_eff}
Consider \hyperref[eq:marginals_short]{2} and substituting $B$ with $B_0 - \frac{d}{B_0}$ and $Q$ with $Q_0 + \frac{d}{Q_0}$ and limits $\lim_{\frac{d}{B_0} \rightarrow 0} \frac{\partial Q}{\partial B}$, $\lim_{\frac{d}{Q_0} \rightarrow 0} \frac{\partial B}{\partial Q}$:

\begin{align*}
    \lim_{\frac{d}{B_0} \rightarrow 0} \frac{\partial Q}{\partial B} &= \lim_{\frac{d}{B_0} \rightarrow 0} \frac{P_B(B_0^2k - (k-1)(B_0 - \frac{d}{B_0})^2)}{P_Q (B_0 - \frac{d}{B_0})^2}\\
    &= \lim_{\frac{d}{B_0} \rightarrow 0} \frac{P_B(B_0 ^ 2k - (k - 1)B_0^2)}{P_Q B_0 ^ 2}\\
    &= \lim_{\frac{d}{B_0} \rightarrow 0} \frac{P_B B_0^2}{P_Q B_0^2}\\
    &= \frac{P_B}{P_Q}
\end{align*}

\begin{align*}
    \lim_{\frac{d}{Q_0} \rightarrow 0} \frac{\partial B}{\partial Q} &= \lim_{\frac{d}{Q_0} \rightarrow 0} \frac{\partial B}{\partial Q} \left(P_Q + \frac{P_Q(P_B B_0 (1 - 2k) + P_Q (Q_0 - (Q_0 + \frac{d}{Q_0})))}{\sqrt{(P_B B_0 (1 - 2k) + P_Q (Q_0 - (Q_0 + \frac{d}{Q_0})))^2 - 4 P_B^2 B_0^2 k (k-1)}} \right) \cdot (2P_B(k-1))^{-1}\\
    &= \lim_{\frac{d}{Q_0} \rightarrow 0} \frac{\partial B}{\partial Q} \left(P_Q + \frac{P_Q(P_B B_0 (1 - 2k) + P_Q (Q_0 - Q_0))}{\sqrt{(P_B B_0 (1 - 2k) + P_Q (Q_0 - Q_0))^2 - 4 P_B^2 B_0^2 k (k-1)}} \right) \cdot (2P_B(k-1))^{-1}\\
    &= \lim_{\frac{d}{Q_0} \rightarrow 0} \frac{\partial B}{\partial Q} \left(P_Q + \frac{P_Q P_B B_0 (1 - 2k)}{\sqrt{(P_B^2 B_0^2 (1 - 2k)^2 - 4 P_B^2 B_0^2 k (k-1)}} \right) \cdot (2P_B(k-1))^{-1}\\
    &= \lim_{\frac{d}{Q_0} \rightarrow 0} \frac{\partial B}{\partial Q} \left(P_Q + \frac{P_Q (1 - 2k)}{\sqrt{((1 - 2k)^2 - 4 k (k-1)}} \right) \cdot (2P_B(k-1))^{-1}\\
    &= \lim_{\frac{d}{Q_0} \rightarrow 0} \frac{\partial B}{\partial Q} \frac{P_Q + P_Q (1 - 2k)}{2P_B(k-1)}\\
    &= \lim_{\frac{d}{Q_0} \rightarrow 0} \frac{2P_Q(1-k)}{2P_B(k-1)}\\
    &= - \frac{P_Q}{P_B}
\end{align*}

These show that as the change in token balance becomes relatively small compared to the pool's total liquidity, the exchange rate approaches the market one.

\subsection{Better Impermanent Loss Recovery with More Liquidity}\label{sec:liquidity_helps_imp_loss}
Consider \hyperref[eq:equilibrium_short]{4} but assume $Q - Q_0 = \frac{d_Q}{Q_0}$ and $B = B_0 - \frac{d_B}{B_0}$. Consider the following limit:

\begin{align*}
    \lim_{\frac{d_Q}{Q_0} \rightarrow 0, \frac{d_B}{B_0} \rightarrow 0} B_{0'} &= \lim_{\frac{d_Q}{Q_0} \rightarrow 0, \frac{d_B}{B_0} \rightarrow 0} \left(B_0 - \frac{d_B}{B_0} \right) + \frac{\left(B_0 - \frac{d_B}{B_0} \right)}{2k} \left(\sqrt{1 + \frac{4k\left(\frac{d}{Q_0} \right)}{\frac{P_{B'}}{P_{Q'}} \left(B_0 - \frac{d_B}{B_0} \right)}} - 1 \right)\\
    &= \lim_{\frac{d_Q}{Q_0} \rightarrow 0, \frac{d_B}{B_0} \rightarrow 0} B_0 + \frac{B_0}{2k} \left(\sqrt{1} - 1 \right)\\
    &= B_0
\end{align*}

Hence, when both tokens' balances deviate little with respect to their starting balances, the recovery procedure identifies equilibrium points which are close to the pool's starting state.

\end{document}